%% file: epistasisEnvFBA_ArXiv.tex
\documentclass[letterpaper]{article}
\usepackage[affil-it]{authblk}

\usepackage{amsmath}
\usepackage{amsthm}

\input{\commonDir preambleCommon}      %
\usepackage[numbers,sort&compress]{natbib}
\graphicspath{{./epistasisEnvFBA/figures/}}

\usepackage{url}
\usepackage{fullpage}
\usepackage{setspace}

\makeatletter
\@addtoreset{algorithm}{section}
\makeatother

\title{Dynamic Epistasis under Varying Environmental Perturbations}

\date{\today}

\begin{document}

\newcounter{LinBrandonCoAuthor}
\author[1]{Brandon Barker\footnote{contributed equally}%
\protect\setcounter{LinBrandonCoAuthor}{\value{footnote}}%
}
\newcommand\LinBrandonCoAuthorMark{\footnotemark[\value{LinBrandonCoAuthor}]}%

\author[2]{Lin Xu\protect\LinBrandonCoAuthorMark}


\author[3,4]{Zhenglong Gu%
  \thanks{Electronic address: \texttt{zg27@cornell.edu}; Corresponding author}}

\affil[1]{Center for Advanced Computing,
    Cornell University, Ithaca, NY, USA.}
\affil[2]{Division of Hematology/Oncology, Department of Pediatrics,
    University of Texas Southwestern Medical Center, Dallas, TX, USA}
\affil[3]{Division of Nutritional Sciences, Cornell University,
  Ithaca, NY, USA.}
\affil[4]{Tri-Institutional Training Program in Computational
  Biology and Medicine, New York, NY, USA.}

\newboolean{thesisStyle}
\setboolean{thesisStyle}{true} 

\maketitle

                                       %
\input{\commonDir documentHeadCommon}  %
                                       %

\begin{abstract}
\epistasisEnviroAbstract
\end{abstract}

\def\suppOrApp{}

\input{\commonDir epistasisEnviroFBA}  %

\section{Supporting Information}

\begin{figure}[!htb]
\centering
\includegraphics[height=0.8\textheight]{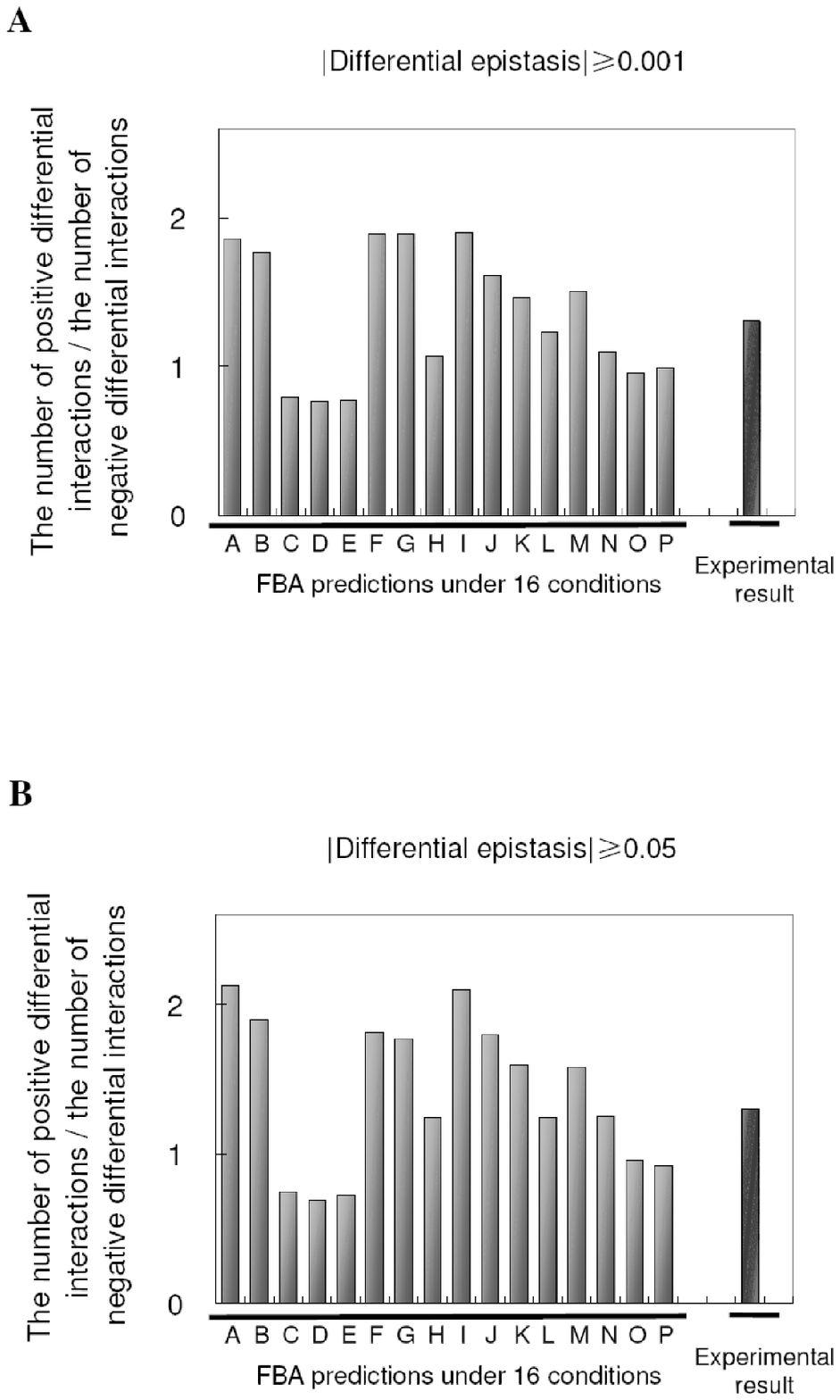}
\caption{\eeFBAfigSOneCap}
\label{fig:eefS1}
\end{figure}

\begin{figure}[!htb]
\centering
\includegraphics[width=0.8\textwidth]{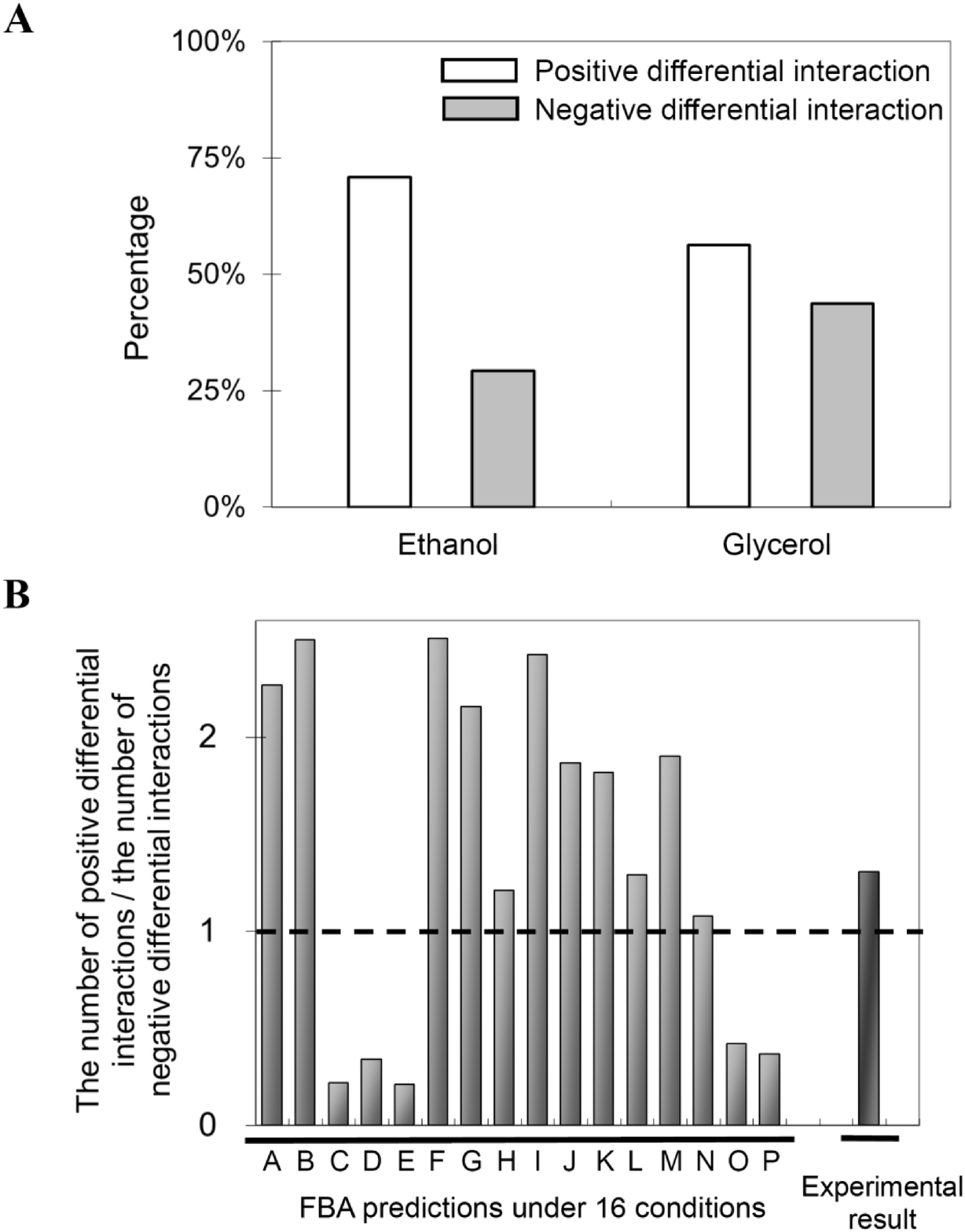}
\caption{\eeFBAfigSTwoCap}
\label{fig:eefS2}
\end{figure}

\begin{figure}[!htb]
\centering
\includegraphics[width=\textwidth]{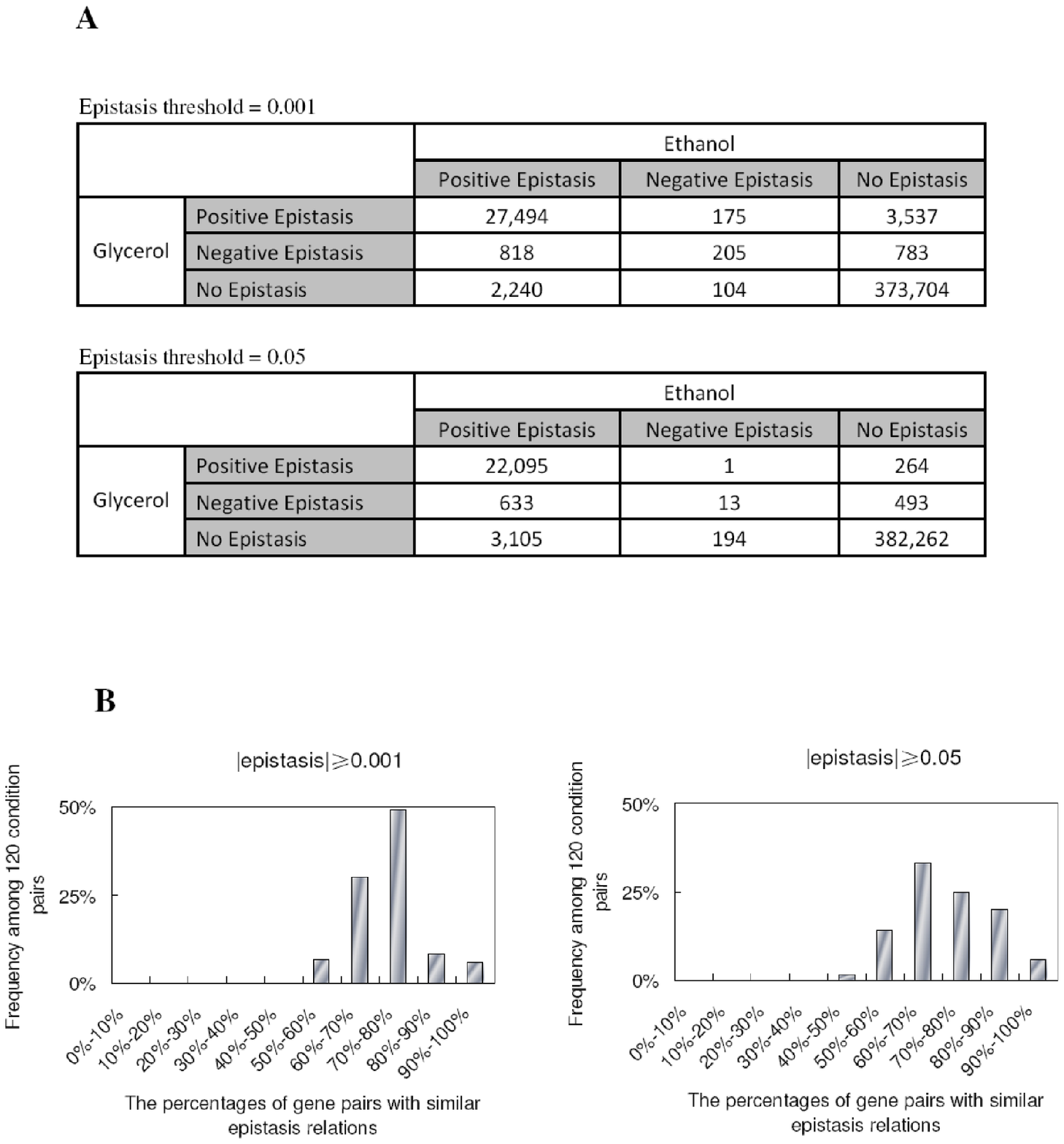}
\caption{\eeFBAfigSThreeCap}
\label{fig:eefS3}
\end{figure}

\begin{figure}[!htb]
\centering
\includegraphics[width=\textwidth]{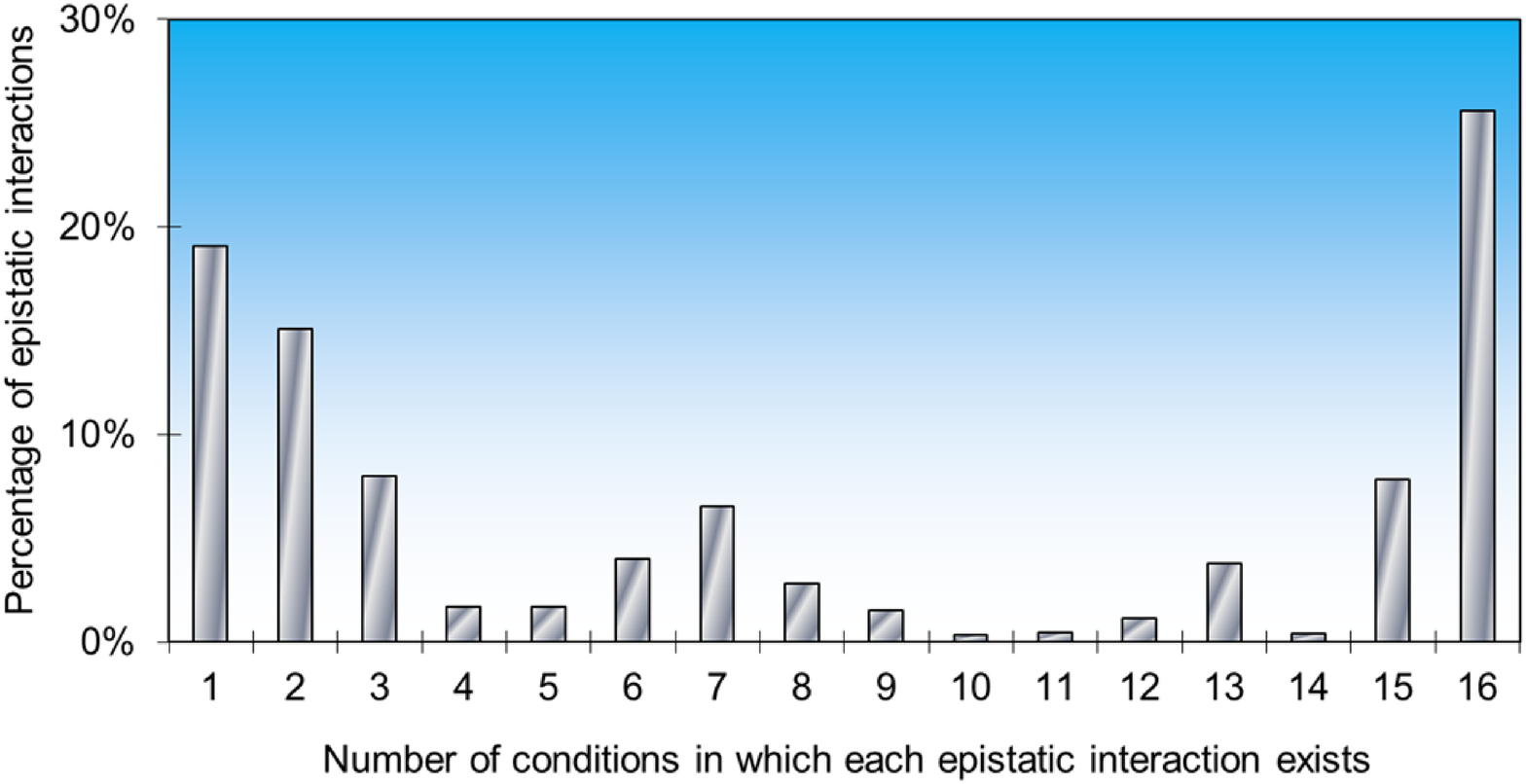}
\caption{\eeFBAfigSFourCap}
\label{fig:eefS4}
\end{figure}

\noindent\textbf{Table S1.} 
\eeFBATabSOneCap

\hspace{1ex}

\noindent\textbf{Table S2.} 
\eeFBATabSTwoCap

\hspace{1ex}

\noindent\textbf{Table S3.} 
\eeFBATabSThreeCap

\hspace{1ex}

\noindent\textbf{Table S4.}
\eeFBATabSFourCap

\hspace{1ex}

\noindent\textbf{Table S5.}
\eeFBATabSFiveCap

\hspace{1ex}

\noindent\textbf{Table S6.}
\eeFBATabSSixCap

\end{document}

%% file: common/preambleCommon.tex
%

\usepackage{caption}

\usepackage{algorithm}
\usepackage{algpseudocode}
\usepackage{float}
\usepackage{ifthen}
\usepackage{soul} 
\usepackage{tikz}
\usepackage[export]{adjustbox}
\usepackage{placeins}

\usepackage{fixltx2e}
\usepackage{upquote,textcomp}

\usepackage{palatino}
\usepackage{subcaption}
\usepackage{txfonts}
\usepackage{array}
\usepackage{calc}
\usepackage{pstricks}
\usepackage{graphicx}
\usepackage{epstopdf}
\usepackage{colortbl}
\usepackage{units}
\usepackage[makeroom]{cancel}

\usetikzlibrary{shapes, arrows}

\makeatletter

\newcommand\floatc@plainthin[2]{
\setbox\@tempboxa\hbox{{\@fs@cfont #1:} #2}
   \hbox to\hsize{\hfil\box\@tempboxa\hfil}\fi}
\newcommand\fs@plainthin{
  \def\@fs@cfont{\rmfamily}\let\@fs@capt\floatc@plainthin
  \def\@fs@pre{}
  \def\@fs@post{\vspace{-2.5em}}
  \def\@fs@mid{\vspace\abovecaptionskip\relax}
  \let\@fs@iftopcapt\iffalse}
\makeatother
\floatstyle{plainthin}
\newfloat{AlgFloat}{h}{lop}

\algnewcommand\algorithmicinput{\textbf{INPUT:}}
\algnewcommand\INPUT{\item[\algorithmicinput]}
\algnewcommand\algorithmicoutput{\textbf{OUTPUT:}}
\algnewcommand\OUTPUT{\item[\algorithmicoutput]}

\makeatletter
\@ifclassloaded{beamer}
{}
{

\newtheoremstyle{break}  
  {\topsep}   
  {\topsep}   
  {\itshape}  
  {-1em}      
  {\bfseries} 
  {}         
  {0pt}  
  {}          
\theoremstyle{break}

}
\makeatother

%% file: common/documentHeadCommon.tex
\newcommand{\captionpage}[2]{
\newpage
\vspace*{\fill}
\begin{center}
\captionof{#1}{#2}
\end{center}
\vspace{\fill}
\newpage
}

\tikzstyle{line} = [draw, very thick, color=black!50, -latex']
\tikzstyle{cloud} = [draw, ellipse,fill=red!20, node distance=2em]
\tikzstyle{decision} = [diamond, draw, fill=blue!20, align=center,
    text badly centered, node distance=6em, inner sep=0pt]
\tikzstyle{block} = [rectangle, draw, 
    text centered, align=center, node distance=4em]
\tikzstyle{iogram} = [trapezium, draw, fill=black!10, 
    trapezium left angle=70, trapezium right angle=-70,
    text centered, align=center, node distance=5em]
\tikzset{
  onslide/.code args={<#1>#2}{\only<#1>{\pgfkeysalso{#2}}}, 
}


\DeclareRobustCommand\suppOrApp{%
  \ifthenelse{\boolean{thesisStyle}}%
    {}%
    {Supplementary}%
}
\DeclareRobustCommand\Fig{%
  \ifthenelse{\boolean{thesisStyle}}%
    {Figure}%
    {Fig.}%
}

\DeclareRobustCommand\Figs{%
  \ifthenelse{\boolean{thesisStyle}}%
    {Figures\:}%
    {Figs.\:}%
}


\ifthenelse{\boolean{thesisStyle}}{%
  \floatstyle{plaintop}
  \restylefloat{table}
}
{}


\newcommand\D{\mathrm{d}}

\newcommand{\ANDw}{\land}
\newcommand{\ORw}{\lor}

\newcommand{\introSameGeneCredit}{%
\ifthenelse{\boolean{thesisStyle}}{%
  \footnote{This chapter is taken from material in \citealt{Shestov2013a}.
Brandon Barker is the primary author of all material found herein.}%
}%
{}%
}


\newcommand{\epiSameGeneCredit}{%
\ifthenelse{\boolean{thesisStyle}}{%
  \footnote{This chapter is published as \citet{Xu2012}.
Brandon Barker and Lin Xu contributed equally to this work.
It is additionally available in \citet[chapter 4]{Xu2012a}.}%
}%
{}%
}

\newcommand{\epiSameGeneAbstract}{%
Epistasis refers to the phenomenon in which phenotypic consequences
caused by mutation of one gene depend on one or more mutations at
another gene. Epistasis is critical for understanding many genetic and
evolutionary processes, including pathway organization, evolution of
sexual reproduction, mutational load, ploidy, genomic complexity,
speciation, and the origin of life. Nevertheless, current
understandings for the genome-wide distribution of epistasis are
mostly inferred from interactions among one mutant type per gene,
whereas how epistatic interaction partners change dynamically for
different mutant alleles of the same gene is largely unknown. Here we
address this issue by combining predictions from flux balance analysis
and data from a recently published high-throughput experiment. Our
results show that different alleles can epistatically interact with
very different gene sets. Furthermore, between two random mutant
alleles of the same gene, the chance for the allele with more severe
mutational consequence to develop a higher percentage of negative
epistasis than the other allele is 50-70\% in eukaryotic organisms,
but only 20-30\% in bacteria and archaea. We developed a population
genetics model that predicts that the observed distribution for the
sign of epistasis can speed up the process of purging deleterious
mutations in eukaryotic organisms. Our results indicate that epistasis
among genes can be dynamically rewired at the genome level, and call
on future efforts to revisit theories that can integrate epistatic
dynamics among genes in biological systems\epiSameGeneCredit.
}


\newcommand{\epistasisEnviroAbstract}{%
Epistasis describes the phenomenon that mutations at different loci do
not have independent effects with regard to certain
phenotypes. Understanding the global epistatic landscape is vital for
many genetic and evolutionary theories. Current knowledge for
epistatic dynamics under multiple conditions is limited by the
technological difficulties in experimentally screening epistatic
relations among genes. We explored this issue by applying flux balance
analysis to simulate epistatic landscapes under various
environmental perturbations. Specifically, we looked at gene-gene
epistatic interactions, where the mutations were assumed to occur in
different genes. We predicted that epistasis tends to become more
positive from glucose-abundant to nutrient-limiting conditions,
indicating that selection might be less effective in removing
deleterious mutations in the latter. We also observed a stable core of
epistatic interactions in all tested conditions, as well as many
epistatic interactions unique to each condition. Interestingly, genes
in the stable epistatic interaction network are directly linked to
most other genes whereas genes with condition-specific epistasis form
a scale-free network. Furthermore, genes with stable epistasis tend to
have similar evolutionary rates, whereas this co-evolving relationship
does not hold for genes with condition-specific epistasis. Our
findings provide a novel genome-wide picture about epistatic dynamics
under environmental perturbations.
}

\newcommand{\epistasisEnviroAuthorSummary}{%
Epistasis, often referred to as genetic interactions, occur when
mutational effects of genes depend on each other. Aside from often
times complicating the way in which the phenotype of an organism
relates to its genotype, epistatic interactions (or epistases) are
essential to several important theories in biology, especially in
evolution. Due to the difficulty in experimentally assessing epistasis
across an entire genome, we employed mathematical modeling of the
metabolic network of baker's yeast to comprehensively simulate genetic
interactions for virtually all known metabolic genes in the
organism. We performed comprehensive simulations in 17 different
environments, which differ by their nutrients. We characterized a
trend that occurs in genetic interactions when yeast is transferred
from a glucose-abundant environment to other environments. We also
found that both the set of genetic interactions present in all
conditions and the set of interactions present in a single environment
are fairly large sets with highly different connectivity. Furthermore,
the set present in all conditions tends to consist of gene pairs with
similar evolutionary rates.
}


\newcommand{\falconAbstractMotivation}{%
A major theme in constraint-based modeling is unifying 
experimental data, such as biochemical information about the reactions
that can occur in a system or the composition and localization of enzyme
complexes, with high-throughput data including expression data,
metabolomics, or DNA sequencing. The desired result is to increase
 predictive capability and improve our understanding of metabolism.
 The approach typically employed when only gene (or protein) intensities
are available is the creation of tissue-specific models, which reduces
the available reactions in an organism model, and does not provide an
objective function for the estimation of fluxes.
}

\newcommand{\falconAbstractResults}{%
We develop a method, flux assignment with LAD (least absolute
deviation) convex objectives and normalization (FALCON),
 that employs metabolic network reconstructions along with expression
data to estimate fluxes. In order to use such a method, accurate
measures of enzyme complex abundance are needed, so we first
present an algorithm that addresses quantification of complex
abundance. Our extensions to prior techniques include the
capability to work with large models and significantly improved
run-time performance even for smaller models, an improved analysis of
enzyme complex formation, the ability to handle large enzyme
complex rules that may incorporate multiple isoforms, and either
maintained or significantly improved correlation with experimentally
measured fluxes.
}

\newcommand{\falconAbstractAvail}{%
FALCON has been implemented in MATLAB and ATS, and can be downloaded
from: \url{https://github.com/bbarker/FALCON}. ATS is not required to
compile the software, as intermediate C source code is available. 
FALCON requires use of the COBRA Toolbox, also implemented in MATLAB.
}


\newcommand{\epiBeneMutAbstract}{%
Existing literature has only dealt with the simulation of strictly
deleterious mutants rather than beneficial mutants in the
constraint-based modeling literature, which is chiefly due to existing
studies optimizing the fitness function, which leaves no room for
improvement. In this study, we develop a constraint-based approach
that can simulate beneficial, neutral, and deleterious mutations.  We
show that this simulation technique can be useful for understanding
adaptive trajectories, and we develop a software library for the
analysis of evolutionary paths.  Our mechanistic model can reproduce
the distribution of
epistases between beneficial mutations that was observed in a data set
and a population genetic model fit to the same data set, showing that
our model behaves appropriately in this conext and may be a useful
tool for further evolutionary analyses. Finally, in experimental data
sets and in our simulations, slightly beneficial mutations are much
more likely to have positive (synergistic) epistasis with other
beneficial mutations, making their likelihood of becoming fixed higher
than would be expected without considering epistatic effects.
}

\captionsetup{labelfont=bf}




%% file: common/epistasisEnviroFBA.tex
\newcommand{\eeFBAfigOneCap}{%
More positive differential epistases under environmental
perturbations. (\textbf{A}) Heat maps describe the global dynamics of
differential epistasis from abundant-glucose medium to ethanol (left
panel) and glycerol (right panel) conditions. Only gene pairs with
$\left|\D\epsilon\right| \geq 0.01$ in either condition are included
in the figure. Different colors represent differential epistasis
values as indicated by the color bar at the bottom. The differential
epistasis values are assigned to be 0.1 (or -0.1) in the heat-maps
when it is greater than 0.1 (or less than -0.1). It is noteworthy to
point out that the epistasis patterns are indeed very different
between the two conditions (\Fig~\ref{fig:eef2}A). (\textbf{B})
Percentage of positive and negative differential epistases under
ethanol and glycerol conditions. (\textbf{C}) Ratio of positive to
negative differential epistases
in each simulated condition. The result from a high-throughput
experiment is also shown. The letters A-P represent acetaldehyde,
acetate, adenosine 3',5'-bisphosphate, adenosyl methionine, adenosine,
alanine, allantoin, arginine, ethanol, glutamate, glutamine, glycerol,
low glucose, phosphate, trehalose, and xanthosine, respectively.
}

\newcommand{\eeFBAfigTwoCap}{%
Epistasis dynamics between environmental perturbations. (\textbf{A}) Number of
gene pairs with various epistatic relationships between ethanol and
glycerol growth conditions. (\textbf{B}) The distribution for the percentages
of gene pairs with similar epistasis relation between any 2 of 16
conditions. The frequency is derived from the 120 pairs of
environments simulated in this study.
}

\newcommand{\eeFBAfigThreeCap}{%
The global distribution of epistatic relations under simulated
conditions. (\textbf{A}) Distribution for the number of conditions in which
each epistatic interaction exists. Note that about 28\% of
epistatic relations are extremely stable (the very right bar) and
about 24\% are extremely dynamic (the very left bar). (\textbf{B}) Fraction
of three types of epistatic relations in each of the 16 environmental
perturbations, as indicated by the color bar to the right. The numbers
in the brackets represent the number of conditions in which each
epistatic interaction exists, as indicated in (\textbf{A}). The letters A-P
represent the simulated conditions as indicated in \Fig~\ref{fig:eef1}.
}

\newcommand{\eeFBAfigFourCap}{%
Network properties for the extremely stable and extremely dynamic
epistatic interactions. (\textbf{A}) Degree distribution for genes in two
epistatic interaction networks. The networks have nodes that
correspond to genes and edges that correspond to epistatic
interactions. (\textbf{B}) Three network parameters (the definition of which
are shown in Methods) for two epistatic interaction networks.
}

\newcommand{\eeFBAfigFiveCap}{%
Co-evolution between genes with epistasis. (\textbf{A}) Average evolutionary
rate differences between gene pairs with FBA-predicted epistasis
(green), extremely dynamic epistasis (blue) and extremely stable
epistasis (red) are highlighted by three arrows, respectively. The
random simulations with the same number of gene pairs as each of the
three groups were repeated 10,000 times and the frequency
distributions are shown (marked by the same colors as the
corresponding arrows, respectively). (\textbf{B}) The evolutionary rates for
genes that are involved in extremely stable and extremely dynamic
epistasis, respectively. The error bars represent standard errors.
}

\newcommand{\eeFBAfigSOneCap}{%
More positive differential epistases under environmental perturbations
for different thresholds of differential epistasis
($\left|\D\epsilon\right| \geq 0.001$, \textbf{A}) and
($\left|\D\epsilon\right| \geq 0.05$, \textbf{B}). Ratio of positive
to negative differential epistases in each simulated condition are
shown. The letters A-P represent acetaldehyde, acetate, adenosine
3',5'-bisphosphate, adenosyl methionine, adenosine, alanine,
allantoin, arginine, ethanol, glutamate, glutamine, glycerol, low
glucose, phosphate, trehalose, and xanthosine, respectively.
}

\newcommand{\eeFBAfigSTwoCap}{%
Analogous to \Fig~\ref{fig:eef1}B-C, but using a maximum growth rate for each
condition, where the maximum is constrained to be no higher than the
high-glucose growth rate. (\textbf{A}) Percentage of positive and
negative differential epistases under ethanol and glycerol
conditions. (\textbf{B}) Ratio of positive to negative differential
epistases in each simulated condition. The result from a
high-throughput experiment is also shown. The letters A-P represent
acetaldehyde, acetate, adenosine 3',5'-bisphosphate, adenosyl
methionine, adenosine, alanine, allantoin, arginine, ethanol,
glutamate, glutamine, glycerol, low glucose, phosphate, trehalose, and
xanthosine, respectively. Note that in (B), low glucose has the same
growth rate as high-glucose, but has different epistatic interactions
since we still use the high-oxygen uptake level associated with the
low glucose condition.
}

\newcommand{\eeFBAfigSThreeCap}{%
Epistasis dynamics between environmental perturbations under different
epistasis definition. (\textbf{A}) Number of gene pairs with various
epistatic relationships between ethanol and glycerol growth conditions
under a lower ($\left|\epsilon\right| \geq 0.001$) and a higher
($\left|\epsilon\right| \geq 0.05$) epistasis threshold. (\textbf{B})
The distribution for the percentages of gene pairs with similar
epistasis relations between any 2 of 16 conditions under a lower
($\left|\epsilon\right| \geq 0.001$) and a higher
($\left|\epsilon\right| \geq 0.05$) epistasis threshold.
}

\newcommand{\eeFBAfigSFourCap}{%
Analogous to \Fig~\ref{fig:eef3}A, but using a maximum growth rate for each
condition, where the maximum is constrained to be no higher than the
high-glucose growth rate. Distribution for the number of conditions in
which each epistatic interaction exists. Note that about 26\% of
epistatic relations are extremely stable (the very right bar) and
about 19\% are extremely dynamic (the very left bar).
}

\newcommand{\eeFBATabSOneCap}{%
Wild-type growth rates used in the maximal growth rate simulations
used for \Figs \ref{fig:eefS2} and \ref{fig:eefS4}.
}

\newcommand{\eeFBATabSTwoCap}{%
Condition-specific epistases and sign-epistases prevalence in the
iMM904 yeast model.
}
\newcommand{\eeFBATabSThreeCap}{%
GO term enrichment analysis results for differential epistasis in
transition to ethanol.
}

\newcommand{\eeFBATabSFourCap}{%
Properties of simulated systems that correlate with the ratio of
positive to negative differential epistases.
}

\newcommand{\eeFBATabSFiveCap}{%
List of epistatic interactions for the extremely stable, dynamic, and
intermediate epistasis networks.
}

\newcommand{\eeFBATabSSixCap}{%
Table of network parameters for stable, dynamic, and intermediate
epistasis.
}

\section{Author Summary} 
\epistasisEnviroAuthorSummary

\section{Introduction}

Epistasis refers to the phenomenon wherein mutations of two genes can
modify each other's phenotypic outcomes. It can be positive
(alleviating), or negative (aggravating), when a combination of
deleterious mutations shows a fitness value that is higher, or lower,
than expectation, respectively. For example, a mutation that hampers a
pathway's function may allow for other mutations in the same pathway
without a fitness consequence, resulting in positive
epistasis. Conversely, genes or pathways with redundant functions can
give rise to negative epistasis. It is well established that epistasis
is important for the evolution of sex \citep{Kondrashov1982,
Azevedo2006, Otto2007}, speciation \citep{Presgraves2007}, mutational
load \citep{Hansen2001}, ploidy \citep{Musso2008}, genetic
architecture of growth traits \citep{Xu2011}, genetic drift
\citep{Perez-Figueroa2009}, genomic complexity \citep{Sanjuan2008},
and drug resistance \citep{Trindade2009}. As biological systems in
nature have to face multiple genetic
and environmental perturbations, understanding the global landscape
and dynamics of epistasis under these perturbations remains an
important issue in the evolutionary field. In an earlier study, we
addressed genome-wide epistasis dynamics under various genetic
perturbations \citep{Xu2012}. In this study, we will investigate the impact of
environmental perturbations on global epistasis dynamics.

How epistatic interactions among genes change in different
environments has been intensively studied in various model
organisms, including \textit{E. coli} \citep{Remold2001,
Kishony2003, Cooper2005}, \textit{S. cerevisiae} \citep{Korona1999,
Szafraniec2001, Jasnos2008}, \textit{C. elegans}
\citep{Vassilieva2000, Baer2006} and \textit{D. melanogaster}
\citep{Yang2001, Fry2002, AletheaD.Wang2009}. The results of these
studies, however, are very controversial. While some studies observed
increasing positive epistasis under harsh conditions
\citep{Kishony2003, Jasnos2008, Yang2001}, others have opposite
findings \citep{Cooper2005, Korona1999, Szafraniec2001, Vassilieva2000,
Baer2006, Fry2002, AletheaD.Wang2009, Young2009}. Even within the same
species, different experimental studies might have conflicting
conclusions (e.g. \citet{Kishony2003, Cooper2005}). One possible
reason for the above controversy could have
originated from the fact that most studies only looked at the
epistasis dynamics based on a small number of genes, where the
properties cannot be generalized to the entire organism.

The main obstacle to exploring global epistatic dynamics under a
variety of environments is the difficulty of applying high-throughput
experimental platforms. To explore epistasis on a genomic scale, a
number of technologies have been developed to systematically map
genetic interaction networks, such as synthetic genetic array (SGA)
\citep{Tong2004, Costanzo2010}, diploid-based synthetic lethality
analysis with microarrays (dSLAM) \citep{Pan2004, Pan2006}, synthetic
dosage-suppression and lethality screen \citep{Measday2002,
Measday2005, Sopko2006} and epistatic miniarray profiles (EMAP)
\citep{Collins2007, Fiedler2009, Kornmann2009}. A key issue for all
these experimental studies is that these epistatic networks have been
constructed only under normal laboratory
conditions. However, cells in nature are constantly bombarded by various
external environmental stresses. Epistasis dynamics under these
perturbations cannot be predicted based on a single laboratory
condition. Few studies have constructed epistatic networks for
multiple environments. A recent study that has only
constructed epistatic networks for a group of genes with specific
functions under one normal and one harsh condition already requires a
large amount of effort \citep{Bandyopadhyay2011}. Consequently,
genome-scale epistasis landscapes under a variety of environmental
perturbations remain largely uncharacterized.

Here we explored this issue by using Flux Balance Analysis (FBA) to
simulate epistasis dynamics among genes under multiple environmental
perturbations. FBA can provide reliable predictions by 
optimizating a presumed objective function,
commonly growth maximization in microbes, subject to the known reactions and
constraints of a metabolic network \citep{Edwards2001, Shlomi2005, Becker2007, Feist2008,
Smallbone2009a, Orth2010}. Using this platform, a previous study has
investigated synthetic lethal interactions (one type of negative
epistasis) under multiple environmental perturbations and showed the
plasticity of epistatic interactions in the metabolic networks
\citep{Harrison2007}. Here we examined both positive and negative
epistasis using FBA,
and were able to show that, on a genome scale, epistatic interactions
tend to become more positive in nutrient-limiting conditions relative
to abundant-glucose media. In addition, while a large proportion of
epistatic interactions can be rewired dynamically under varying
environments, there is a set of epistatic interactions that are
stable across all tested environments. We also discovered different
network and evolutionary properties for genes with stable and dynamic
epistatic interactions. Implications of our findings were discussed.

\section{Results}

\subsection{FBA modeling and simulated growth conditions}

We applied the yeast \textit{S. cerevisiae} metabolic reconstruction
iMM904 \citep{Mo2009} to examine the dynamics of epistasis under various
environmental perturbations. The reconstruction has 904 metabolic
genes that are associated with 1,412 metabolic reactions. We conducted
FBA simulations under an abundant-glucose condition and 16
nutrient-limiting conditions. In 15 of these conditions, the carbon source
(abundant glucose) was replaced by one of the following: acetaldehyde,
acetate, adenosine 3',5'-bisphosphate, adenosyl methionine, adenosine,
alanine, allantoin, arginine, ethanol, glutamate, glutamine, glycerol,
low glucose, trehalose, and xanthosine, respectively. These conditions
represent a wide variety of nutrient and energy sources: nucleosides,
amino acids, sugars, alcohols, etc. Additionally, we looked at
abundant glucose under limited phosphorus availability.

To ensure that all environments have the same growth
rates in the following analyses, we restricted the carbon source or
phosphorous uptake levels for each of the 16 environmental
perturbations such that only 20\% of the high-glucose growth rate was
attained. This was chosen because it has been shown that metabolism
is directly linked to growth and similar growth rates often induce
similar metabolic pathways \citep{Jakubowska2012}. It is therefore
important to use a fixed growth rate among different conditions to
control for the relationship between growth rates and the overall
metabolic activity so as not to induce a growth-rate specific
effect. The 20\% high-glucose level was chosen because some media
types do not support high growth rates, regardless of the abundance of
the nutrient source. Specifically, the acetate condition had the
lowest wild-type growth rate (38.5\% of the wild-type glucose growth
rate) when unrestricted carbon uptake was permitted. In order to allow
flexibility with adding more conditions in future studies while
simultaneously not allowing extremely low growth rates that may be
possible to model but are unlikely to persist in natural environments,
we chose a growth rate of approximately half of this minimal wild-type
growth rate as the the wild-type growth rate for all conditions.

In order to estimate epistasis between genes, we
created a mutation for each gene in each condition that restricted the
flux to be 50\% of the wild-type flux found by geometric FBA for all
reactions associated with the mutant gene \citep{Xu2012}. Epistatic relations
between any two genes were calculated under each condition. We also
tested our core findings allowing maximum growth in each condition
(Table S1) and the general trends in our results remained similar, as
described in the following.

\subsection{More positive differential epistases from rich media to
nutrient-limiting conditions}

\begin{figure}[!htb]
\centering
\includegraphics[height=0.75\textheight]{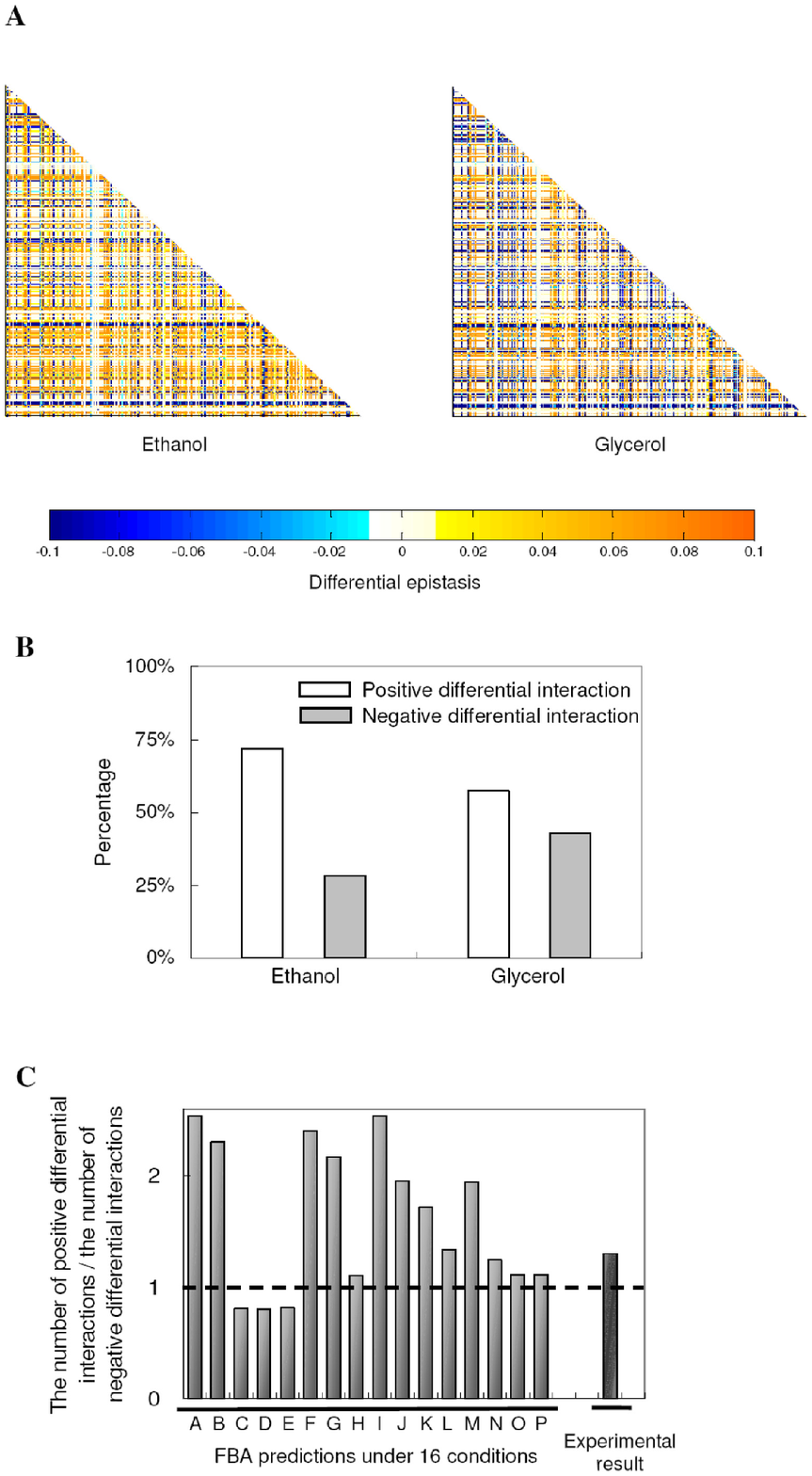}
\caption{\eeFBAfigOneCap}
\label{fig:eef1}
\end{figure}

To directly address how the sign and magnitude of epistases change
under nutrient-limiting conditions, we calculated differential
epistasis ($\D\epsilon$), which is defined as the epistatic change from
abundant-glucose media to the nutrient-limiting condition for each
gene pair in each growth condition. A gene pair with positive (or
negative) differential interaction under an environmental perturbation
is defined as the gene pair having increasing (or decreasing)
epistasis values from the abundant-glucose media to that
condition. \Fig~\ref{fig:eef1}A depicts the
distribution of differential
epistases in two growth conditions (ethanol and glycerol) as an
example. Only genes with $\left|\D\epsilon\right| \geq 0.01$ in at least
one of the two conditions are included in this figure. As quantified
in Table S2, there are 6.1\% and 5.5\% of all gene pairs with
$\left|\D\epsilon\right| \geq 0.01$ from abundant-glucose media to ethanol
and glycerol growth conditions, respectively. Among them, a large
number of gene pairs even change their sign of epistasis (Table
S2). Simulations in other conditions show similar effects (Table S2),
indicating that epistatic relationships among genes can be very
dynamic between abundant-glucose media and nutrient-limiting
conditions.



We further investigated the sign of differential epistasis from
abundant-glucose to nutrient-limiting conditions. As shown in \Fig~\ref{fig:eef1}A, 
we observed more yellow dots (positive differential epistasis)
than blue dots (negative differential epistasis) in both
panels. Indeed, as quantified in \Fig~\ref{fig:eef1}B, 72\% and 57\% of
differential epistases are positive in ethanol and glycerol
conditions, respectively. We further explored all 16 nutrient-limiting
conditions and the results are shown in \Fig~\ref{fig:eef1}C. In most of our
simulated conditions (13/16), there are significantly more positive
differential epistases than negative differential epistases (Binomial
test, P $< 10^{-5}$ for each of the 13 conditions), indicating that
epistasis tends to become more positive in nutrient-limiting
conditions. This conclusion does not depend on the criteria we used to
define differential epistasis (\Fig~\ref{fig:eefS1}).

A recent high-throughput experiment measured epistatic relations
between roughly 80,000 gene pairs with and without perturbation by a
DNA-damaging agent (methyl methanesulfonate, MMS). The study
represents the most comprehensive experimental study so far to explore
epistatic dynamics from a rich medium to a harsh condition
\citep{Bandyopadhyay2011}. Interestingly, the authors also found more
positive differential
epistases than negative differential epistases, which is consistent
with our general observation (\Fig~\ref{fig:eef1}C). We further allowed maximum
growth in each condition and the general trends in our results
remained similar (\Fig~\ref{fig:eefS2}).

We found that differential epistasis had functional importance after
performing both Gene Ontology (GO) and Kyoto Encyclopedia of Genes and
Genomes (KEGG) enrichment analyses to compare genes with positive and
negative differential epistases through the glucose-abundant to
ethanol transition. We chose the ethanol condition as an example
because it is one of the most widely used conditions for the baker's
yeast. We observed that 38 GO terms and 8 KEGG pathways
are enriched for positive differential epistasis, while 18 GO terms
and 1 KEGG pathways are enriched for negative differential epistasis
(Table S3). More importantly, we found positive and negative
differential epistases uniquely contribute to different aspects of
ethanol and energy metabolism. For example, positive differential
epistasis is enriched in monohydric alcohol metabolic processes,
oxidoreductase activity acting on aldehyde group donors, the TCA
cycle, and pyruvate metabolism, while negative differential epistasis
is enriched in ethanol metabolic processes and various amino acid
terms and pathways, indicating the functional importance of
differential epistasis (Table S3). This is consistent with
experimental results that show differential epistatic interactions,
rather than static epistatic interactions, are functionally related to
the response of interest \citep{Bandyopadhyay2011}.

Several system properties were found to correlate with the ratio of
the number of positive to negative differential epistases (Table
S4). A strong correlation exists between the number of essential genes
in a given condition and the ratio of positive to negative
differential epistasis on transition from high glucose to that
condition ($\rho = 0.8392$, P $=4.8107 \times 10^{-5}$), which was a better predictor
than the number of non-zero fluxes in the wild-type vector for that
environment ($\rho = 0.5115$, P = 0.0428). Another strong predictor for
positive differential epistasis was the mean relative fitness of
single mutants in the new environment ($\rho = -0.8029$, P $= 2.7427 \times 10^{-4}$);
this anticorrelation suggests that a propensity for a lower single
mutant fitness can cause a shift towards positive epistasis.

\subsection{Dynamic epistasis between nutrient-limiting conditions}

\begin{figure}[!htb]
\centering
\includegraphics[width=0.9\textwidth]{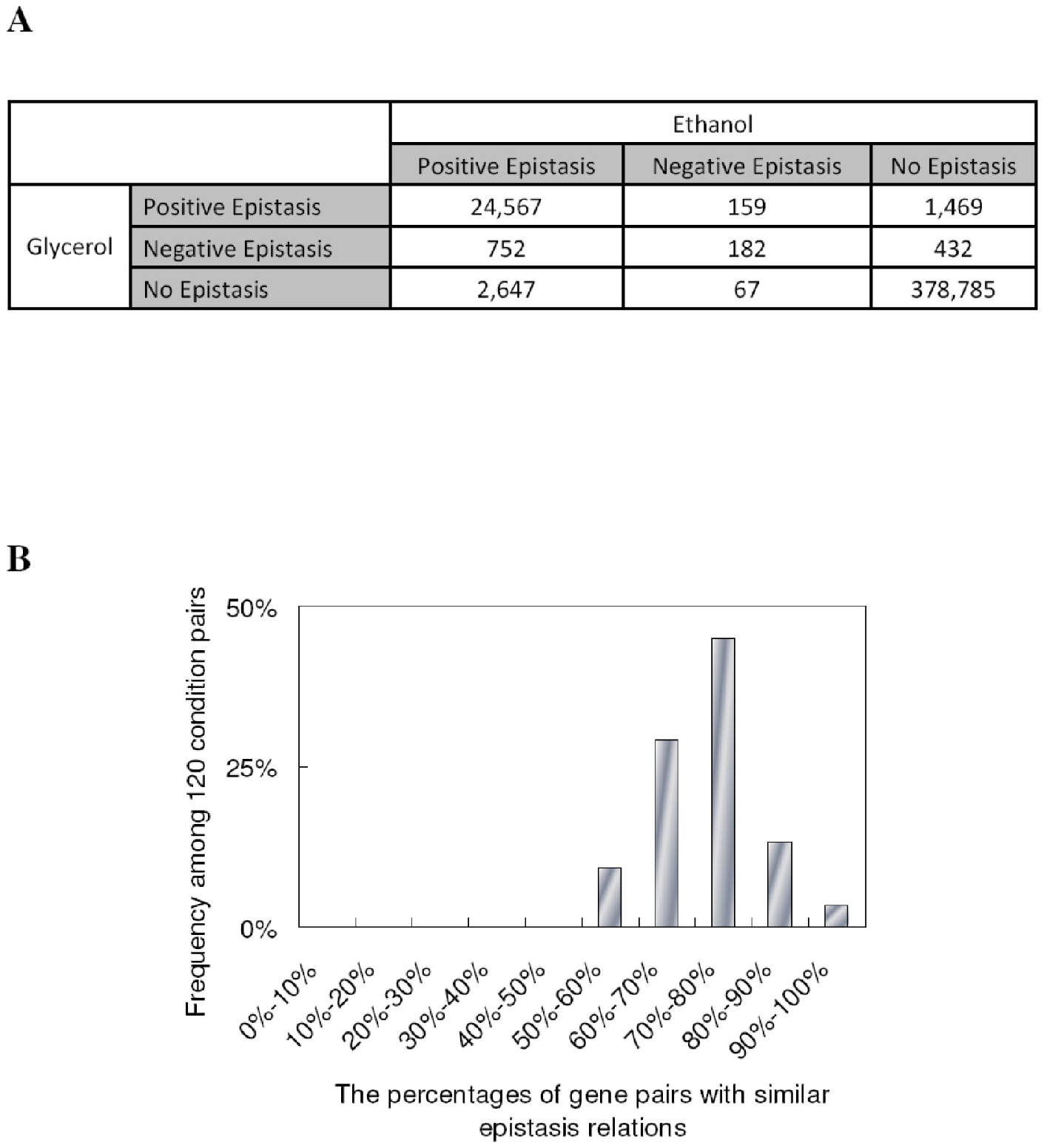}
\caption{\eeFBAfigTwoCap}
\label{fig:eef2}
\end{figure}

\Fig~\ref{fig:eef1} explored the epistasis dynamics from abundant-glucose media
to nutrient-limiting conditions. As biological systems in nature
constantly face perturbations to the environment, it is
interesting to investigate the epistasis dynamics among
nutrient-limiting conditions. To achieve this aim, we first explored
the epistatic relationship between the same gene pairs in ethanol and
glycerol environments. \Fig~\ref{fig:eef2}A lists the number of gene pairs that have various
epistatic relationships. It is noteworthy to point out that,
consistent with previously published results, there are significantly
more positive epistases than negative epistases between genes in either
condition \citep{He2010}.


If two genes have the same sign of epistasis and
$\left|\epsilon\right| \geq 0.01$ in both conditions, they are defined
as having a similar epistatic
relationship in these two conditions. To quantify epistatic dynamics
between ethanol and glycerol growth conditions, we defined the
percentage of gene pairs with similar epistatic relations to be the
number of gene pairs with similar epistasis relations shared in these
two conditions (overlap) divided by the number of gene pairs with
epistasis in either condition (union). Our results show that 79\% of
gene pairs have similar epistasis relations between these two
conditions. \Fig~\ref{fig:eef2}B shows the distribution for the percentages of
gene pairs with similar epistasis relations between any 2 of 16
conditions, demonstrating a variable degree of epistatic similarity
between any two conditions.  This conclusion still holds when we used
different criteria to define epistatic relationships between genes
(\Fig~\ref{fig:eefS3}).

\begin{figure}[!htb]
\centering
\includegraphics[width=0.85\textwidth]{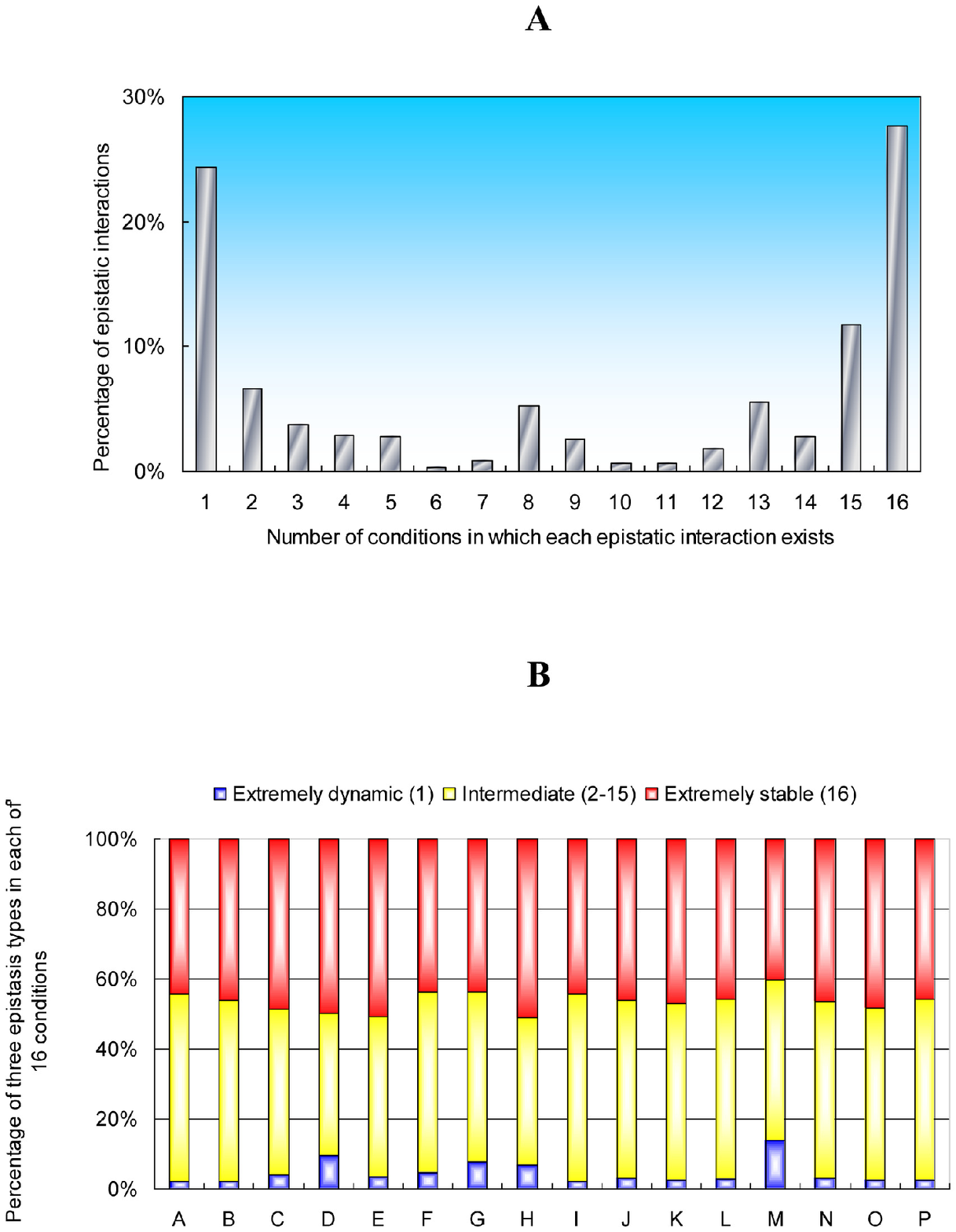}
\caption{\eeFBAfigThreeCap}
\label{fig:eef3}
\end{figure}

To understand the global distribution of all epistatic relations, we
considered 16 conditions together and calculated the fraction of
epistatic interactions existing in 1, 2, 3, \ldots, 15, and 16
conditions, respectively. As shown in \Fig~\ref{fig:eef3}A, we found that there is a
U frequency distribution for the number of growth conditions in which
a specific epistatic interaction is observed. This means that
approximately 52\% of these interactions are either condition-specific
(24\%; termed dynamic) or predicted to exist in all conditions (28\%;
termed stable), and about 48\% is intermediate (exists in multiple but not
all 16 conditions). An analogous result was obtained previously, but
only for synthetic lethal interactions \citep{Harrison2007}. We also changed the
growth assumption and allowed maximum growth in each condition and
reanalyzed the global distribution of all epistatic relations. The U
frequency distribution for the number of growth conditions in which a
specific epistatic interaction is observed remained similar
(\Fig~\ref{fig:eefS4}). Based on the result in \Fig~\ref{fig:eef3}A,
we further calculated the ratio of these three types of epistatic
relations in each of the 16 environmental perturbations. As shown in
\Fig~\ref{fig:eef3}B, we found that in
each environment, about 40-60\% of epistatic interactions are stable
and that each environment also has many private epistases among genes.



\subsection{Different network properties for stable and dynamic epistasis}

\begin{figure}[!htb]
\centering
\includegraphics[width=\textwidth]{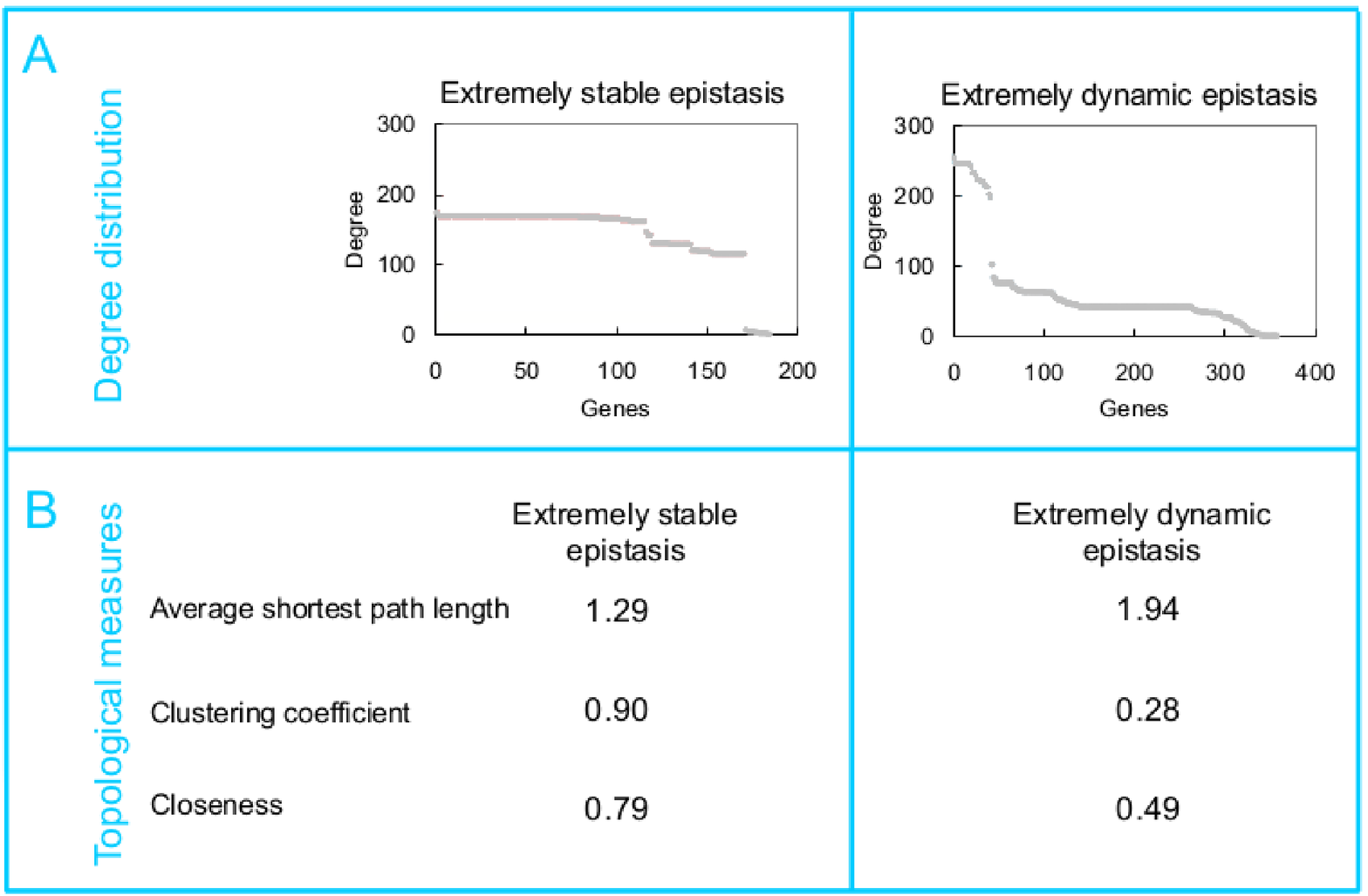}
\caption{\eeFBAfigTwoCap}
\label{fig:eef4}
\end{figure}

Analysis on network properties can reveal various organization
principles (e.g. frequency of occurrence, centrality) for epistasis
networks \citep{Tong2004, Costanzo2010} and therefore provide valuable
information to further
distinguish stable and dynamic epistasis. To achieve this aim, we
compared networks formed by extremely stable and extremely dynamic
epistasis among genes and asked whether they have distinct network
properties. The degree distributions for both types of epistasis are
shown in \Fig~\ref{fig:eef4}A. Interestingly, extremely stable epistatic
interactions form an exponential network architecture, which is
homogeneous, meaning that most nodes have a very similar number of
links (\Fig~\ref{fig:eef4}A, left panel). In contrast, the extremely dynamic
epistatic interactions give rise to a scale-free network topology,
which is heterogeneous, meaning that the majority of nodes have few
links but a small number of hubs have a large number of links
(\Fig~\ref{fig:eef4}A, right panel).


In addition, we calculated three network parameters to compare these
two types of epistatic interactions. We found that the network formed
by extremely stable epistases has a smaller shortest path length, a
larger clustering coefficient and larger closeness than the network
formed by extremely dynamic epistases (\Fig~\ref{fig:eef4}B). These results are
consistent with the scenario that genes with extremely stable
epistasis are directly linked to most other genes and form an
exponential network topology, while genes with extremely dynamic
epistasis form a scale-free network. Our results also show that the
network induced by intermediate epistases have intermediate values of
these parameters compared to that of extremely stable and extremely
dynamic epistasis networks (Tables S5 and S6).

\subsection{Co-evolution of genes with epistatic interaction}

\begin{figure}[!htb]
\centering
\includegraphics[height=0.75\textheight]{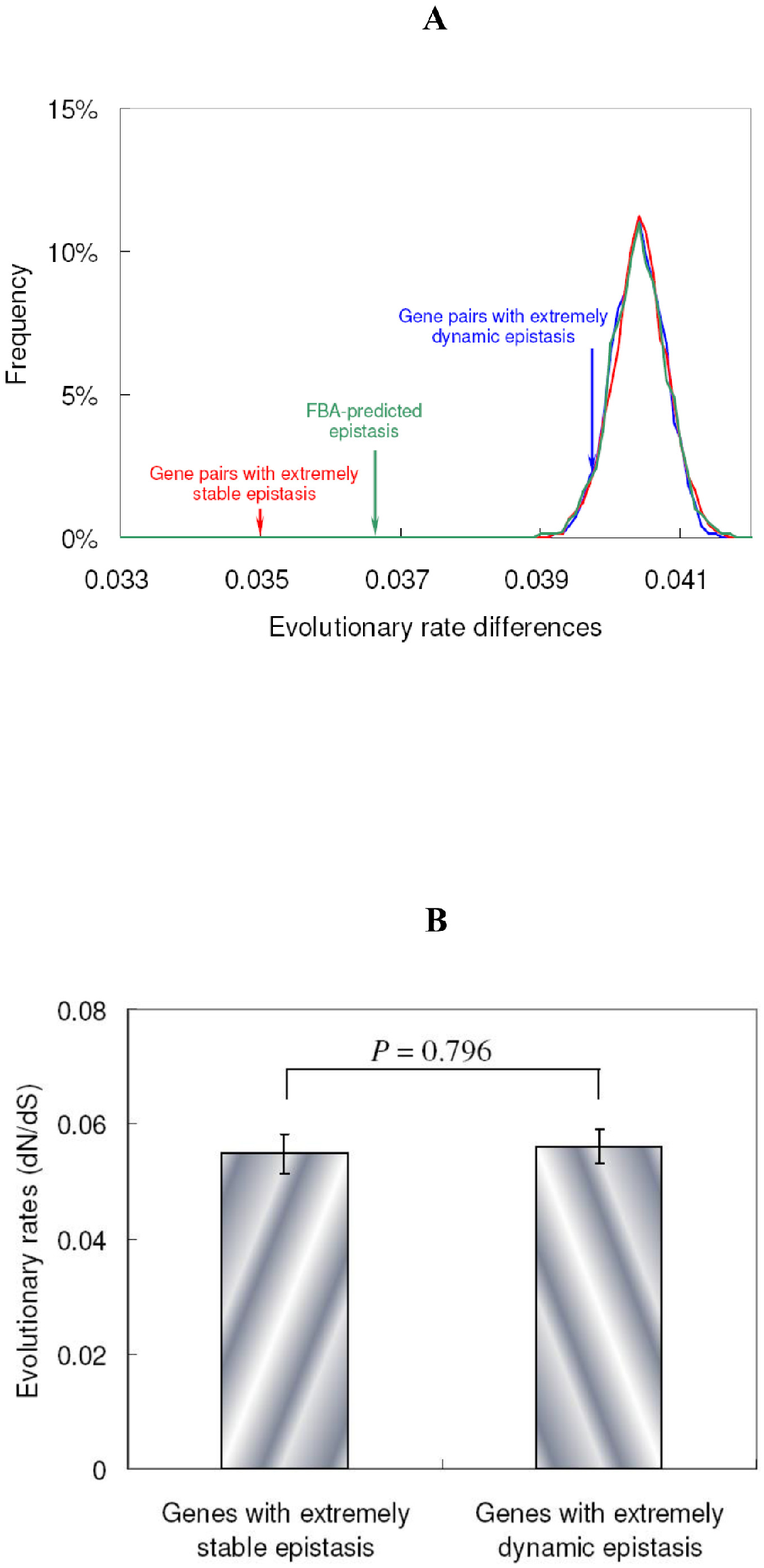}
\caption{\eeFBAfigFiveCap}
\label{fig:eef5}
\end{figure}

Gene pairs with epistasis identified in real experiments usually show
similar evolutionary rates. To investigate whether two genes with
predicted epistasis also tend to co-evolve, we calculated the
evolutionary rate differences between two genes with epistasis from
FBA modeling (\Fig~\ref{fig:eef5}A). Evolutionary rates
($\nicefrac{d_\textnormal{N}}{d_\textnormal{S}}$) based on orthologous
gene sets
from four yeast species of the genus Saccharomyces were downloaded
from a commonly used reference dataset \citep{Wall2005}. Simulations
based on the same number of gene pairs with FBA-predicted epistasis
were conducted to estimate the evolutionary rate differences for any
two randomly selected genes. As shown in \Fig~\ref{fig:eef5}A, the gene pairs
with FBA-predicted epistatic interactions tend to have more similar
evolutionary rates than random expectation (P $< 10^{-4}$).



In \Fig~\ref{fig:eef4} we observed unique network properties for extremely stable
and extremely dynamic epistatic interactions. We further investigate
the co-evolution between genes with these two types of epistatic
relationships. As shown in \Fig~\ref{fig:eef5}A, genes with extremely stable
epistasis tend to co-evolve (P $< 10^{-4}$), while the difference between
genes with extremely dynamic epistasis and random expectation becomes
much smaller (P = 0.06). The evolutionary rate difference between gene
pairs with extremely stable and extremely dynamic epistasis is also
significant (\textit{t}-test, P $= 8\times 10^{-6}$). This difference is not caused by
genes that are involved in extremely stable or extremely dynamic
epistasis, because these two groups of genes do not have significantly
different evolutionary rates (\textit{t}-test, P = 0.796, \Fig~\ref{fig:eef5}B).

\section{Discussion}

\subsection{Natural selection in nutrient-limiting conditions}

Whether a genetic mutation has a fitness consequence depends on other
sites, a phenomenon called epistasis (see \citet{Lehner2011} for a
recent review on
molecular mechanisms). Positive epistasis alleviates the total harm
when multiple deleterious mutations combine together and thus reduces
the effectiveness of natural selection in removing these deleterious
mutations, whereas negative epistasis plays the opposite role by
increasing the efficiency of purging deleterious mutations by natural
selection. Results from this study present an initial glimpse over
environment-induced epistasis dynamics at the genome scale. Using
differential epistasis from abundant-glucose to nutrient-limiting
conditions, our results show that epistasis between specific genes can
become more positive or more negative in nutrient-limiting conditions,
which is consistent with previous findings in small scale studies
\citep{Remold2001, Kishony2003, Cooper2005, Korona1999, Szafraniec2001,
Jasnos2008, Vassilieva2000, Baer2006, Yang2001, Fry2002,
AletheaD.Wang2009, Young2009}. However, we showed that, at the genome
scale, epistasis is
more positive in nutrient-limiting conditions. Interestingly, our
simulation results are consistent with a recent genome-wide study
between laboratory and harsh growth conditions
\citep{Bandyopadhyay2011}. How epistasis affects selection in harsh
conditions has been controversial \citep{Agrawal2010}. Our
results provide the genome-wide evidence arguing that selection might
be less effective in removing deleterious mutations in harsh
conditions, which could be one of the underlying reasons for a recent
observation that stimulation of a stress response can reduce mutation
penetrance in \textit{Caenorhabditis elegans} \citep{Casanueva2012}.

\subsection{Network properties and evolutionary patterns for stable
and dynamic epistasis}

Our results indicate that epistasis could be extremely stable or
dynamic among various environmental perturbations, which is consistent
with a previous FBA study investigating synthetic lethal relations
among non-essential genes \citep{Harrison2007}. The inclusion of
essential genes in our study allows for investigation on many
important metabolic
pathways that were not previously analyzed. Nevertheless, the
distribution of epistasis among multiple environments (\Fig~\ref{fig:eef3}A)
remains largely unchanged from the previous study \citep{Harrison2007}
even when essential genes are included.

We also found that stable and dynamic epistatic relationships show
totally different network properties and evolutionary patterns, which
might provide new biological and evolutionary insights. The gene pairs
with stable epistases tend to co-evolve with each other. In addition,
from the biological pathway perspective, the smaller shortest path
length and larger closeness values in the stable epistasis network
both imply that genes with stable epistases tend to be functionally
associated with a large number of neighbors to form a condensed
functional network, different from genes in the dynamic epistasis
network that are loosely connected. Furthermore, the large clustering
coefficient in the stable epistasis network also supports the idea
that genes with stable epistasis interactions form a network core in
the whole epistasis network. Combined with observations in \Fig~\ref{fig:eef5},
this core module of epistasis in the metabolic network might represent
stable functional associations between genes that are essential for
important biological functions and evolutionarily conserved even under
different environmental perturbations. The lack of co-evolutionary
pattern and scale-free network properties for the dynamic epistasis
network, however, might represent unstable functional associations
between genes, which may only be responsible for unique functions
under specific conditions.

\subsection{Implications and significance for exploring stable and
dynamic epistasis}

Our prediction about stable and dynamic epistasis could have important
functional applications. A recent study showed that the synthetic
lethal (negative epistasis) relationship between fumarate hydratase
and haem oxygenase can be employed successfully to identify an in
vitro drug target in hereditary leiomyomatosis and renal-cell cancer
(HLRCC) cells \citep{Frezza2011}. Exploring both dynamic and stable
epistasis could be useful in this context; stable epistatic
interactions may be
important for drug target detection in cancer or other pathogens,
whereas it may sometimes be necessary to exploit dynamic epistatic
relationships, possibly induced by treatment with an external
perturbation.

Furthermore, rational evolutionary design techniques such as OptKnock
\citep{Burgard2003} and OptGene \citep{Patil2005} attempt to find
which knockouts will enable a reaction of interest to be coupled with
growth (i.e. have positive
epistasis with growth-associated genes in a specific
environment). However, these techniques do not take into account
epistasis dynamics across different environments. In this study, we
have found that epistatic relations can be highly dynamic under
various environmental perturbations, which raises the possibility to
improve these techniques by considering epistasis dynamics in future
studies. Research on using compensatory perturbations to reach desired
network states is ongoing \citep{Cornelius2011}.

\subsection{Caveats and future directions} Though we show several
novel insights into how varying environments can influence epistasis,
several caveats should be addressed. First, the FBA modeling used in
this study, which was proven to have great predictive power and has
been successfully employed in addressing numerous research problems
\citep{Edwards2001, Shlomi2005, Smallbone2009a}, 
only includes metabolic genes. Second, even though FBA
offers the most comprehensive simulation method for studying
epistasis, there are many improvements that can be made in order to
capture the empirically observed set of epistatic interactions
\citep{Covert2001}. For example, integrating transcriptional
regulation and physical interactions into this framework could improve
the current methods in predicting epistasis and other evolutionary
processes \citep{Szappanos2011}. Related to
this point, FBA as used herein only considers the steady state and
does not take into account any dynamics or initial conditions, and
would necessarily miss any epistatic interactions that are due to
dynamics in the system, such as changing concentrations; dynamic FBA
(which is part of rFBA) might be a solution, but would likely require
about a minimum of two orders of magnitude increase in computation
time \citep{Covert2001, Varma1994}. Recent work
on new objective functions targeting metabolite turnover rather than
flux per se has also proven successful in recovering many epistases
that were previously not found with FBA \citep{Brochado2012}.

Third, in order to understand the impact of environmental
perturbations on epistasis, we used a reductive approach and only
considered one mutation type per gene to simulate the global epistatic
landscape in 16 environments. There are countless environments in
nature. Furthermore, different mutations in the same gene and the
interactions between genes and environment can likely have an even
more complex impact on the epistasis dynamics. While it would be ideal
to simulate a larger variety of environments for multiple
mutations of the same gene, the computational cost is a limiting
factor. Our previous study showed that different mutants of the same
gene could have very dynamic epistatic interaction partners in a
single environment \citep{Xu2012}. In this study, we chose to use one
mutation per gene as we are focusing on addressing how different
environments could affect gene epistasis dynamics. Nevertheless, in
order to see how sensitive our results were, we performed the analysis
for our core results by simulating 16 environments using different
growth assumption, where the organisms are allowed to have
unrestricted uptake of the limiting nutrient to obtain the maximum
growth in that condition. We found the major trends in our results are
largely unchanged (\Figs \ref{fig:eefS2} and \ref{fig:eefS4}; Table S1).

Keeping these issues in mind, our analysis uncovered several prominent
features of epistatic interactions under a variety of environmental
perturbations, and call on future effort to confirm these simulation
results using high-throughput experimental platforms. More
importantly, the enrichment of stable and dynamic epistasis provides a
new perspective to understand how biological systems may rewire
epistasis in nature.

\section{Methods}

Scripts for generating and analyzing the data can be found in the
source code repository located at
\url{https://github.com/bbarker/COBRAscripts/}. Scripts and
documentation specific to this paper are located in the subdirectory
\texttt{MyProjects/EnvironmentalEpistasisFBA}.

\subsection{Flux Balance Analysis}

Flux Balance Analysis attempts to tackle issues inherent in other
methods of metabolic modeling, such as the need to measure a large
number of parameters, slow speed of simulation, and dependence on
initial conditions \citep{Orth2010, Schellenberger2011a}. 
Other than needing a fairly complete
understanding of the reactions present in an organism, the only
measurements required to perform a genome-scale metabolic simulation
are those for determining biomass constitution or a gene expression
profile \citep{Shlomi2005, Mo2009}. Strictly speaking, FBA is a particular type of
constraint based modeling (CBM). Constraint based modeling frames the
stoichiometry that describe the reactions present in an organism as a
matrix equation with indeterminates (reaction fluxes) subject to
constraints \citep{Smallbone2009a, Mo2009}. The optimization problem
is described as follows:

\begin{equation}
\centering
\begin{tabular}{rl}
maximize   & $\mathbf{c}^T\mathbf{v}$                                     \\
subject to & $\mathbf{S v} = \frac{\D{\mathbf{x}}}{\D{t}} = \mathbf{0}$   \\
           & $\mathbf{v}_{lb} \preceq \mathbf{v} \preceq \mathbf{v}_{ub}$ \\
\end{tabular}
\end{equation}

$\mathbf{S}$ is a matrix, in which rows and columns correspond to
cellular metabolites and reactions in the reconstructed network
respectively. $\mathbf{v}$ is the reaction flux with upper and lower
bounds $\mathbf{v}_{ub}$ and $\mathbf{v}_{lb}$
respectively. Multiplying the stoichiometric matrix $\mathbf{S}$ by
the flux vector v equals the concentration change over time
($\frac{\D{\mathbf{x}}}{\D{t}}$). At steady state, the flux through
each reaction is given by $\mathbf{Sv} = \mathbf{0}$. Further details
on the underlying methods can be found in the literature 
\citep{Xu2012, Smallbone2009a, He2010}.

The fluxes of mutations employed in this analysis were restricted to
be 50\% of the wild-type fluxes found for growth rate maximization by
geometric FBA \citep{He2010}. To find new conditions with a specified carbon
source or other limiting nutrient that achieves 20\% of the
high-glucose growth rate, we can solve a linear program for the
minimization of the limiting nutrient uptake while requiring the
growth rate to be equal to 20\% of the abundant-glucose
growth-rate. For maximum growth rate conditions (Table S1, \Figs \ref{fig:eefS2}
and \ref{fig:eefS4}), we allowed unrestricted uptake of the limiting nutrient to
obtain the maximum growth in that condition, up to the point where it
would reach the high-glucose growth rate. Mutations affecting protein
complexes and pleiotropic genes are handled by uniform restriction
across enzymes as described before \citep{Xu2012}.

\subsection{Definition of epistasis}

In each gene mutant pair, the epistasis value is calculated based on
the equation: $\epsilon = W_{xy} - W_xW_y$, in which $W_{xy}$ is the
fitness of an organism with two mutations in genes X and Y, whereas
$W_{x}$ or $W_{y}$ refers to the fitness of the organism with mutation
only at gene X or Y respectively. Each fitness listed previously is
calculated relative to the wild-type fitness. Absolute fitness values
are determined by the value of the biomass maximization objective
present in the model. Finally, a confidence threshold
($\left|\epsilon\right| \geq 0.01$) was applied to generate epistatic
interactions \citep{Xu2012, Costanzo2010, He2010}. We also conducted
analyses based on a different threshold for epistasis and the general
conclusions still hold in our analysis (\Figs \ref{fig:eefS1} and
\ref{fig:eefS3}).

\subsection{Evolutionary rates and network parameters}

Evolutionary rates of \textit{S. cerevisiae} genes were downloaded
from supplementary materials of \citet{Wall2005}, in which orthologs were
defined by four complete genomes of \textit{Saccharomyces} species
(\textit{Saccharomyces cerevisiae}, \textit{Saccharomyces paradoxus},
\textit{Saccharomyces mikatae} and \textit{Saccharomyces bayanus}) and
evolutionary rates at synonymous and nonsynonymous sites were
calculated based on a four-way yeast species alignment for
\textit{S. cerevisiae} genes by PAML. For the distributions in
\ref{fig:eef5}A, we randomly sampled gene pairs with the same number
of gene pairs
as in three epistasis networks (epistasis in all 16 conditions,
extremely stable epistasis, and extremely dynamic epistasis,
respectively), and calculated the average evolutionary rate
differences between random gene pairs in each of these three sample
sets. The simulations were repeated 10,000 times for each of the three
groups, which are color coded to correspond to the epistasis networks
of the same size.

Network parameters such as the shortest path length, clustering
coefficient and closeness were calculated using the computer software
Pajek, downloaded from:
\url{http://vlado.fmf.uni-lj.si/pub/networks/pajek}. The shortest path
length between two genes in a network reflects the overall network
interconnectedness; the smaller the average shortest path length is,
the higher chance that genes in this network could interact with the
other genes. The clustering coefficient of a network is a measurement
of the degree to which nodes in a network tend to cluster together;
the larger the average clustering coefficient is, the more closely the
genes are connected, forming modules. The closeness of a network
measures the centrality of nodes within a network; nodes that occur on
shortest paths with other nodes have higher closeness than those that
do not \citep{Barabasi2004}.

\section{Acknowledgments}

We firstly thank Tim Connallon for extremely helpful discussion
related to epistasis. Discussions from Dr. Huifeng Jiang, Dr. Chris
Myers, and Mr. Kaixiong Ye are also very much appreciated. This work
was supported by the startup fund from Cornell University, NSF
DEB-0949556 and NIH 1R01AI085286 awarded to Z.G, and the 
Tri-Institutional fellowship, awarded to B.B.

%% file: epistasisEnvFBA_ArXiv.bbl
\clearpage

\begin{thebibliography}{58}
\providecommand{\natexlab}[1]{#1}
\providecommand{\url}[1]{\texttt{#1}}
\expandafter\ifx\csname urlstyle\endcsname\relax
  \providecommand{\doi}[1]{doi: #1}\else
  \providecommand{\doi}{doi: \begingroup \urlstyle{rm}\Url}\fi

\bibitem[Kondrashov(1982)]{Kondrashov1982}
Alexey~S Kondrashov.
\newblock {Selection against harmful mutations in large sexual and asexual
  populations}.
\newblock \emph{Genet. Res. (Camb).}, 40\penalty0 (03):\penalty0 325--332,
  1982.

\bibitem[Azevedo et~al.(2006)Azevedo, Lohaus, Srinivasan, Dang, and
  Burch]{Azevedo2006}
Ricardo B~R Azevedo, Rolf Lohaus, Suraj Srinivasan, Kristen~K Dang, and
  Christina~L Burch.
\newblock {Sexual reproduction selects for robustness and negative epistasis in
  artificial gene networks.}
\newblock \emph{Nature}, 440\penalty0 (7080):\penalty0 87--90, March 2006.
\newblock ISSN 1476-4687.
\newblock \doi{10.1038/nature04488}.
\newblock URL \url{http://www.ncbi.nlm.nih.gov/pubmed/16511495}.

\bibitem[Otto(2007)]{Otto2007}
Sarah~P Otto.
\newblock {Unravelling the evolutionary advantage of sex: a commentary on
  'Mutation-selection balance and the evolutionary advantage of sex and
  recombination' by Brian Charlesworth.}
\newblock \emph{Genet. Res.}, 89\penalty0 (5-6):\penalty0 447--9, December
  2007.
\newblock ISSN 1469-5073.
\newblock \doi{10.1017/S001667230800966X}.
\newblock URL \url{http://www.ncbi.nlm.nih.gov/pubmed/18976534}.

\bibitem[Presgraves(2007)]{Presgraves2007}
Daven~C Presgraves.
\newblock {Speciation Genetics: Epistasis, Conflict and the Origin of Species}.
\newblock \emph{Curr. Biol.}, 17\penalty0 (4):\penalty0 R125--R127, July 2007.
\newblock \doi{10.1016/j.cub.2006.12.030}.
\newblock URL
  \url{http://www.cell.com/current-biology/abstract/S0960-9822(06)02664-9}.

\bibitem[Hansen and Wagner(2001)]{Hansen2001}
Thomas~F Hansen and G\"{u}nter~P Wagner.
\newblock {Epistasis and the Mutation Load: A Measurement-Theoretical
  Approach}.
\newblock \emph{Genet.}, 158\penalty0 (1):\penalty0 477--485, May 2001.
\newblock URL \url{http://www.genetics.org/content/158/1/477.abstract}.

\bibitem[Musso et~al.(2008)Musso, Costanzo, Huangfu, Smith, Paw, {San Luis},
  Boone, Giaever, Nislow, Emili, and Zhang]{Musso2008}
Gabriel Musso, Michael Costanzo, ManQin Huangfu, Andrew~M Smith, Jadine Paw,
  Bryan-Joseph {San Luis}, Charles Boone, Guri Giaever, Corey Nislow, Andrew
  Emili, and Zhaolei Zhang.
\newblock {The extensive and condition-dependent nature of epistasis among
  whole-genome duplicates in yeast}.
\newblock \emph{Genome Res.}, 18\penalty0 (7):\penalty0 1092--1099, July 2008.
\newblock \doi{10.1101/gr.076174.108}.
\newblock URL \url{http://genome.cshlp.org/content/18/7/1092.abstract}.

\bibitem[Xu et~al.(2011)Xu, Jiang, Chen, and Gu]{Xu2011}
Lin Xu, Huifeng Jiang, Hong Chen, and Zhenglong Gu.
\newblock {Genetic Architecture of Growth Traits Revealed by Global Epistatic
  Interactions}.
\newblock \emph{Genome Biol. Evol.}, 3:\penalty0 909--914, January 2011.
\newblock \doi{10.1093/gbe/evr065}.
\newblock URL \url{http://gbe.oxfordjournals.org/content/3/909.abstract}.

\bibitem[P\'{e}rez-Figueroa et~al.(2009)P\'{e}rez-Figueroa, Caballero,
  Garc\'{\i}a-Dorado, and L\'{o}pez-Fanjul]{Perez-Figueroa2009}
Andr\'{e}s P\'{e}rez-Figueroa, Armando Caballero, Aurora Garc\'{\i}a-Dorado,
  and Carlos L\'{o}pez-Fanjul.
\newblock {The Action of Purifying Selection, Mutation and Drift on Fitness
  Epistatic Systems}.
\newblock \emph{Genet.}, 183\penalty0 (1):\penalty0 299--313, September 2009.
\newblock \doi{10.1534/genetics.109.104893}.
\newblock URL \url{http://www.genetics.org/content/183/1/299.abstract}.

\bibitem[Sanju\'{a}n and Nebot(2008)]{Sanjuan2008}
Rafael Sanju\'{a}n and Miguel~R. Nebot.
\newblock {A network model for the correlation between epistasis and genomic
  complexity}.
\newblock \emph{PLoS One}, 3, 2008.
\newblock ISSN 19326203.
\newblock \doi{10.1371/journal.pone.0002663}.

\bibitem[Trindade et~al.(2009)Trindade, Sousa, Xavier, Dionisio, Ferreira, and
  Gordo]{Trindade2009}
Sandra Trindade, Ana Sousa, Karina~Bivar Xavier, Francisco Dionisio,
  Miguel~Godinho Ferreira, and Isabel Gordo.
\newblock {Positive Epistasis Drives the Acquisition of Multidrug Resistance}.
\newblock \emph{PLoS Genet}, 5\penalty0 (7):\penalty0 e1000578, July 2009.
\newblock URL \url{http://dx.doi.org/10.1371\%2Fjournal.pgen.1000578}.

\bibitem[Xu et~al.(2012)Xu, Barker, and Gu]{Xu2012}
Lin Xu, Brandon Barker, and Zhenglong Gu.
\newblock {Dynamic epistasis for different alleles of the same gene.}
\newblock \emph{Proc. Natl. Acad. Sci. U. S. A.}, June 2012.
\newblock ISSN 1091-6490.
\newblock \doi{10.1073/pnas.1121507109}.
\newblock URL \url{http://www.ncbi.nlm.nih.gov/pubmed/22689976}.

\bibitem[Remold and Lenski(2001)]{Remold2001}
Susanna~K Remold and Richard~E Lenski.
\newblock {Contribution of individual random mutations to
  genotype-by-environment interactions in Escherichia coli}.
\newblock \emph{Proc. Natl. Acad. Sci.}, 98\penalty0 (20):\penalty0
  11388--11393, September 2001.
\newblock \doi{10.1073/pnas.201140198}.
\newblock URL \url{http://www.pnas.org/content/98/20/11388.abstract}.

\bibitem[Kishony and Leibler(2003)]{Kishony2003}
Roy Kishony and Stanislas Leibler.
\newblock {Environmental stresses can alleviate the average deleterious effect
  of mutations}.
\newblock \emph{J. Biol.}, 2\penalty0 (2):\penalty0 14, 2003.
\newblock ISSN 1475-4924.
\newblock URL \url{http://jbiol.com/content/2/2/14}.

\bibitem[Cooper et~al.(2005)Cooper, Lenski, and Elena]{Cooper2005}
Tim~F Cooper, Richard~E Lenski, and Santiago~F Elena.
\newblock {Parasites and mutational load: an experimental test of a pluralistic
  theory for the evolution of sex.}
\newblock \emph{Proc. Biol. Sci.}, 272:\penalty0 311--317, 2005.
\newblock ISSN 0962-8452.
\newblock \doi{10.1098/rspb.2004.2975}.

\bibitem[Korona(1999)]{Korona1999}
Ryszard Korona.
\newblock {Genetic Load of the Yeast Saccharomyces cerevisiae under Diverse
  Environmental Conditions}.
\newblock \emph{Evolution (N. Y).}, 53\penalty0 (6):\penalty0 1966--1971,
  December 1999.
\newblock ISSN 00143820.
\newblock \doi{10.2307/2640455}.
\newblock URL \url{http://www.jstor.org/stable/2640455}.

\bibitem[Szafraniec et~al.(2001)Szafraniec, Borts, and Korona]{Szafraniec2001}
Krzysztof Szafraniec, Rhona~H Borts, and Ryszard Korona.
\newblock {Environmental stress and mutational load in diploid strains of the
  yeast Saccharomyces cerevisiae}.
\newblock \emph{Proc. Natl. Acad. Sci.}, 98\penalty0 (3):\penalty0 1107--1112,
  January 2001.
\newblock \doi{10.1073/pnas.98.3.1107}.
\newblock URL \url{http://www.pnas.org/content/98/3/1107.abstract}.

\bibitem[Jasnos et~al.(2008)Jasnos, Tomala, Paczesniak, and Korona]{Jasnos2008}
Lukasz Jasnos, Katarzyna Tomala, Dorota Paczesniak, and Ryszard Korona.
\newblock {Interactions Between Stressful Environment and Gene Deletions
  Alleviate the Expected Average Loss of Fitness in Yeast}.
\newblock \emph{Genet.}, 178\penalty0 (4):\penalty0 2105--2111, April 2008.
\newblock \doi{10.1534/genetics.107.084533}.
\newblock URL \url{http://www.genetics.org/content/178/4/2105.abstract}.

\bibitem[Vassilieva et~al.(2000)Vassilieva, Hook, and Lynch]{Vassilieva2000}
Larissa~L Vassilieva, Aaron~M Hook, and Michael Lynch.
\newblock {THE FITNESS EFFECTS OF SPONTANEOUS MUTATIONS IN CAENORHABDITIS
  ELEGANS}.
\newblock \emph{Evolution (N. Y).}, 54\penalty0 (4):\penalty0 1234--1246,
  August 2000.
\newblock ISSN 1558-5646.
\newblock \doi{10.1111/j.0014-3820.2000.tb00557.x}.
\newblock URL \url{http://dx.doi.org/10.1111/j.0014-3820.2000.tb00557.x}.

\bibitem[Baer et~al.(2006)Baer, Phillips, Ostrow, Avalos, Blanton, Boggs,
  Keller, Levy, and Mezerhane]{Baer2006}
Charles~F Baer, Naomi Phillips, Dejerianne Ostrow, Ari\'{a}n Avalos, Dustin
  Blanton, Ashley Boggs, Thomas Keller, Laura Levy, and Edward Mezerhane.
\newblock {Cumulative Effects of Spontaneous Mutations for Fitness in
  Caenorhabditis: Role of Genotype, Environment and Stress}.
\newblock \emph{Genet.}, 174\penalty0 (3):\penalty0 1387--1395, November 2006.
\newblock \doi{10.1534/genetics.106.061200}.
\newblock URL \url{http://www.genetics.org/content/174/3/1387.abstract}.

\bibitem[Yang et~al.(2001)Yang, Tanikawa, {Van Voorhies}, Silva, and
  Kondrashov]{Yang2001}
Hsiao-Pei Yang, Ana~Y Tanikawa, Wayne~A {Van Voorhies}, Joana~C Silva, and
  Alexey~S Kondrashov.
\newblock {Whole-Genome Effects of Ethyl Methanesulfonate-Induced Mutation on
  Nine Quantitative Traits in Outbred Drosophila melanogaster}.
\newblock \emph{Genet.}, 157\penalty0 (3):\penalty0 1257--1265, March 2001.
\newblock URL \url{http://www.genetics.org/content/157/3/1257.abstract}.

\bibitem[Fry and Heinsohn(2002)]{Fry2002}
James~D Fry and Stefanie~L Heinsohn.
\newblock {Environment Dependence of Mutational Parameters for Viability in
  Drosophila melanogaster}.
\newblock \emph{Genet.}, 161\penalty0 (3):\penalty0 1155--1167, July 2002.
\newblock URL \url{http://www.genetics.org/content/161/3/1155.abstract}.

\bibitem[Wang et~al.(2009)Wang, Sharp, Spencer, Tedman-Aucoin, and
  Agrawal]{AletheaD.Wang2009}
Alethea~D Wang, Nathaniel~P Sharp, Christine~C Spencer, Katherine
  Tedman-Aucoin, and Aneil~F Agrawal.
\newblock {Selection, Epistasis, and Parent‐of‐Origin Effects on
  Deleterious Mutations across Environments in Drosophila melanogaster.}
\newblock \emph{Am. Nat.}, 174\penalty0 (6):\penalty0 863--874, December 2009.
\newblock ISSN 00030147.
\newblock \doi{10.1086/645088}.
\newblock URL \url{http://www.jstor.org/stable/10.1086/645088}.

\bibitem[Young et~al.(2009)Young, Yourth, and Agrawal]{Young2009}
Jadene~A Young, Christopher~P Yourth, and Aneil~F Agrawal.
\newblock {The effect of pathogens on selection against deleterious mutations
  in Drosophila melanogaster}.
\newblock \emph{J. Evol. Biol.}, 22\penalty0 (10):\penalty0 2125--2129, October
  2009.
\newblock ISSN 1420-9101.
\newblock \doi{10.1111/j.1420-9101.2009.01830.x}.
\newblock URL \url{http://dx.doi.org/10.1111/j.1420-9101.2009.01830.x}.

\bibitem[Tong et~al.(2004)Tong, Lesage, Bader, Ding, Xu, Xin, Young, Berriz,
  Brost, Chang, Chen, Cheng, Chua, Friesen, Goldberg, Haynes, Humphries, He,
  Hussein, Ke, Krogan, Li, Levinson, Lu, M\'{e}nard, Munyana, Parsons, Ryan,
  Tonikian, Roberts, Sdicu, Shapiro, Sheikh, Suter, Wong, Zhang, Zhu, Burd,
  Munro, Sander, Rine, Greenblatt, Peter, Bretscher, Bell, Roth, Brown,
  Andrews, Bussey, and Boone]{Tong2004}
Amy Hin~Yan Tong, Guillaume Lesage, Gary~D Bader, Huiming Ding, Hong Xu,
  Xiaofeng Xin, James Young, Gabriel~F Berriz, Renee~L Brost, Michael Chang,
  YiQun Chen, Xin Cheng, Gordon Chua, Helena Friesen, Debra~S Goldberg,
  Jennifer Haynes, Christine Humphries, Grace He, Shamiza Hussein, Lizhu Ke,
  Nevan Krogan, Zhijian Li, Joshua~N Levinson, Hong Lu, Patrice M\'{e}nard,
  Christella Munyana, Ainslie~B Parsons, Owen Ryan, Raffi Tonikian, Tania
  Roberts, Anne-Marie Sdicu, Jesse Shapiro, Bilal Sheikh, Bernhard Suter,
  Sharyl~L Wong, Lan~V Zhang, Hongwei Zhu, Christopher~G Burd, Sean Munro,
  Chris Sander, Jasper Rine, Jack Greenblatt, Matthias Peter, Anthony
  Bretscher, Graham Bell, Frederick~P Roth, Grant~W Brown, Brenda Andrews,
  Howard Bussey, and Charles Boone.
\newblock {Global Mapping of the Yeast Genetic Interaction Network}.
\newblock \emph{Science (80-. ).}, 303\penalty0 (5659):\penalty0 808--813,
  February 2004.
\newblock \doi{10.1126/science.1091317}.
\newblock URL \url{http://www.sciencemag.org/content/303/5659/808.abstract}.

\bibitem[Costanzo et~al.(2010)Costanzo, Baryshnikova, Bellay, Kim, Spear,
  Sevier, Ding, Koh, Toufighi, Mostafavi, Prinz, {St Onge}, VanderSluis,
  Makhnevych, Vizeacoumar, Alizadeh, Bahr, Brost, Chen, Cokol, Deshpande, Li,
  Lin, Liang, Marback, Paw, {San Luis}, Shuteriqi, Tong, van Dyk, Wallace,
  Whitney, Weirauch, Zhong, Zhu, Houry, Brudno, Ragibizadeh, Papp, P\'{a}l,
  Roth, Giaever, Nislow, Troyanskaya, Bussey, Bader, Gingras, Morris, Kim,
  Kaiser, Myers, Andrews, and Boone]{Costanzo2010}
Michael Costanzo, Anastasia Baryshnikova, Jeremy Bellay, Yungil Kim, Eric~D
  Spear, Carolyn~S Sevier, Huiming Ding, Judice L~Y Koh, Kiana Toufighi, Sara
  Mostafavi, Jeany Prinz, Robert~P {St Onge}, Benjamin VanderSluis, Taras
  Makhnevych, Franco~J Vizeacoumar, Solmaz Alizadeh, Sondra Bahr, Renee~L
  Brost, Yiqun Chen, Murat Cokol, Raamesh Deshpande, Zhijian Li, Zhen-Yuan Lin,
  Wendy Liang, Michaela Marback, Jadine Paw, Bryan-Joseph {San Luis}, Ermira
  Shuteriqi, Amy Hin~Yan Tong, Nydia van Dyk, Iain~M Wallace, Joseph~A Whitney,
  Matthew~T Weirauch, Guoqing Zhong, Hongwei Zhu, Walid~A Houry, Michael
  Brudno, Sasan Ragibizadeh, Bal\'{a}zs Papp, Csaba P\'{a}l, Frederick~P Roth,
  Guri Giaever, Corey Nislow, Olga~G Troyanskaya, Howard Bussey, Gary~D Bader,
  Anne-Claude Gingras, Quaid~D Morris, Philip~M Kim, Chris~A Kaiser, Chad~L
  Myers, Brenda~J Andrews, and Charles Boone.
\newblock {The genetic landscape of a cell.}
\newblock \emph{Science}, 327\penalty0 (5964):\penalty0 425--31, 2010.
\newblock ISSN 1095-9203.
\newblock \doi{10.1126/science.1180823}.
\newblock URL \url{http://www.ncbi.nlm.nih.gov/pubmed/20093466}.

\bibitem[Pan et~al.(2004)Pan, Yuan, Xiang, Wang, Sookhai-Mahadeo, Bader,
  Hieter, Spencer, and Boeke]{Pan2004}
Xuewen Pan, Daniel~S. Yuan, Dong Xiang, Xiaoling Wang, Sharon Sookhai-Mahadeo,
  Joel~S. Bader, Philip Hieter, Forrest Spencer, and Jef~D. Boeke.
\newblock {A robust toolkit for functional profiling of the yeast genome}.
\newblock \emph{Mol. Cell}, 16:\penalty0 487--496, 2004.
\newblock ISSN 10972765.
\newblock \doi{10.1016/j.molcel.2004.09.035}.

\bibitem[Pan et~al.(2006)Pan, Ye, Yuan, Wang, Bader, and Boeke]{Pan2006}
Xuewen Pan, Ping Ye, Daniel~S. Yuan, Xiaoling Wang, Joel~S. Bader, and Jef~D.
  Boeke.
\newblock {A DNA integrity network in the yeast Saccharomyces cerevisiae}.
\newblock \emph{Cell}, 124:\penalty0 1069--1081, 2006.
\newblock ISSN 00928674.
\newblock \doi{10.1016/j.cell.2005.12.036}.

\bibitem[Measday and Hieter(2002)]{Measday2002}
V~Measday and P~Hieter.
\newblock {Synthetic dosage lethality}.
\newblock In \emph{Guid. TO YEAST Genet. Mol. CELL Biol. PT B}, volume 350 of
  \emph{METHODS IN ENZYMOLOGY}, pages 316--326. ACADEMIC PRESS INC, 525 B
  STREET, SUITE 1900, SAN DIEGO, CA 92101-4495 USA, 2002.

\bibitem[Measday et~al.(2005)Measday, Baetz, Guzzo, Yuen, Kwok, Sheikh, Ding,
  Ueta, Hoac, Cheng, Pot, Tong, Yamaguchi-Iwai, Boone, Hieter, and
  Andrews]{Measday2005}
Vivien Measday, Kristin Baetz, Julie Guzzo, Karen Yuen, Teresa Kwok, Bilal
  Sheikh, Huiming Ding, Ryo Ueta, Trinh Hoac, Benjamin Cheng, Isabelle Pot, Amy
  Tong, Yuko Yamaguchi-Iwai, Charles Boone, Phil Hieter, and Brenda Andrews.
\newblock {Systematic yeast synthetic lethal and synthetic dosage lethal
  screens identify genes required for chromosome segregation.}
\newblock \emph{Proc. Natl. Acad. Sci. U. S. A.}, 102:\penalty0 13956--13961,
  2005.
\newblock ISSN 0027-8424.
\newblock \doi{10.1073/pnas.0503504102}.

\bibitem[Sopko et~al.(2006)Sopko, Huang, Preston, Chua, Papp, Kafadar, Snyder,
  Oliver, Cyert, Hughes, Boone, and Andrews]{Sopko2006}
Richelle Sopko, Dongqing Huang, Nicolle Preston, Gordon Chua, Bal\'{a}zs Papp,
  Kimberly Kafadar, Mike Snyder, Stephen~G. Oliver, Martha Cyert, Timothy~R.
  Hughes, Charles Boone, and Brenda Andrews.
\newblock {Mapping pathways and phenotypes by systematic gene overexpression}.
\newblock \emph{Mol. Cell}, 21:\penalty0 319--330, 2006.
\newblock ISSN 10972765.
\newblock \doi{10.1016/j.molcel.2005.12.011}.

\bibitem[Collins et~al.(2007)Collins, Miller, Maas, Roguev, Fillingham, Chu,
  Schuldiner, Gebbia, Recht, Shales, Ding, Xu, Han, Ingvarsdottir, Cheng,
  Andrews, Boone, Berger, Hieter, Zhang, Brown, Ingles, Emili, Allis, Toczyski,
  Weissman, Greenblatt, and Krogan]{Collins2007}
Sean~R Collins, Kyle~M Miller, Nancy~L Maas, Assen Roguev, Jeffrey Fillingham,
  Clement~S Chu, Maya Schuldiner, Marinella Gebbia, Judith Recht, Michael
  Shales, Huiming Ding, Hong Xu, Junhong Han, Kristin Ingvarsdottir, Benjamin
  Cheng, Brenda Andrews, Charles Boone, Shelley~L Berger, Phil Hieter, Zhiguo
  Zhang, Grant~W Brown, C~James Ingles, Andrew Emili, C~David Allis, David~P
  Toczyski, Jonathan~S Weissman, Jack~F Greenblatt, and Nevan~J Krogan.
\newblock {Functional dissection of protein complexes involved in yeast
  chromosome biology using a genetic interaction map.}
\newblock \emph{Nature}, 446:\penalty0 806--810, 2007.
\newblock ISSN 0028-0836.
\newblock \doi{10.1038/nature05649}.

\bibitem[Fiedler et~al.(2009)Fiedler, Braberg, Mehta, Chechik, Cagney,
  Mukherjee, Silva, Shales, Collins, van Wageningen, Kemmeren, Holstege,
  Weissman, Keogh, Koller, Shokat, and Krogan]{Fiedler2009}
Dorothea Fiedler, Hannes Braberg, Monika Mehta, Gal Chechik, Gerard Cagney,
  Paromita Mukherjee, Andrea~C Silva, Michael Shales, Sean~R Collins, Sake van
  Wageningen, Patrick Kemmeren, Frank C~P Holstege, Jonathan~S Weissman,
  Michael-Christopher Keogh, Daphne Koller, Kevan~M Shokat, and Nevan~J Krogan.
\newblock {Functional organization of the S. cerevisiae phosphorylation
  network.}
\newblock \emph{Cell}, 136\penalty0 (5):\penalty0 952--63, 2009.
\newblock ISSN 1097-4172.
\newblock \doi{10.1016/j.cell.2008.12.039}.
\newblock URL \url{http://www.ncbi.nlm.nih.gov/pubmed/19269370}.

\bibitem[Kornmann et~al.(2009)Kornmann, Currie, Collins, Schuldiner, Nunnari,
  Weissman, and Walter]{Kornmann2009}
Beno\^{\i}t Kornmann, Erin Currie, Sean~R Collins, Maya Schuldiner, Jodi
  Nunnari, Jonathan~S Weissman, and Peter Walter.
\newblock {An ER-mitochondria tethering complex revealed by a synthetic biology
  screen.}
\newblock \emph{Science}, 325:\penalty0 477--481, 2009.
\newblock ISSN 0036-8075.
\newblock \doi{10.1126/science.1175088}.

\bibitem[Bandyopadhyay(2011)]{Bandyopadhyay2011}
Sourav Bandyopadhyay.
\newblock {Rewiring of Genetic Networks in Response to DNA Damage}.
\newblock \emph{Science (80-. ).}, 1385\penalty0 (2010), 2011.
\newblock \doi{10.1126/science.1195618}.

\bibitem[Edwards et~al.(2001)Edwards, Ibarra, and Palsson]{Edwards2001}
J~S Edwards, R~U Ibarra, and B~O Palsson.
\newblock {In silico predictions of Escherichia coli metabolic capabilities are
  consistent with experimental data.}
\newblock \emph{Nat. Biotechnol.}, 19:\penalty0 125--130, 2001.
\newblock ISSN 1087-0156.
\newblock \doi{10.1038/84379}.

\bibitem[Shlomi et~al.(2005)Shlomi, Berkman, and Ruppin]{Shlomi2005}
Tomer Shlomi, Omer Berkman, and Eytan Ruppin.
\newblock {Regulatory on/off minimization of metabolic flux}.
\newblock \emph{Proc. Natl. Acad. Sci.}, 102\penalty0 (21):\penalty0
  7695--7700, 2005.

\bibitem[Becker et~al.(2007)Becker, Feist, Mo, Hannum, Palsson, and
  Herrg\aa~rd]{Becker2007}
Scott~A Becker, Adam~M Feist, Monica~L Mo, Gregory Hannum, Bernhard~\O Palsson,
  and Markus~J Herrg\aa~rd.
\newblock {Quantitative prediction of cellular metabolism with constraint-based
  models: the COBRA Toolbox.}
\newblock \emph{Nat. Protoc.}, 2\penalty0 (3):\penalty0 727--38, 2007.
\newblock ISSN 1750-2799.
\newblock \doi{10.1038/nprot.2007.99}.
\newblock URL \url{http://www.ncbi.nlm.nih.gov/pubmed/17406635}.

\bibitem[Feist and Palsson(2008)]{Feist2008}
Adam~M Feist and Bernhard~\O Palsson.
\newblock {The growing scope of applications of genome-scale metabolic
  reconstructions using Escherichia coli.}
\newblock \emph{Nat. Biotechnol.}, 26:\penalty0 659--667, 2008.
\newblock ISSN 1087-0156.
\newblock \doi{10.1038/nbt1401}.

\bibitem[Smallbone and Simeonidis(2009)]{Smallbone2009a}
Kieran Smallbone and Evangelos Simeonidis.
\newblock {Flux balance analysis: a geometric perspective.}
\newblock \emph{J. Theor. Biol.}, 258\penalty0 (2):\penalty0 311--5, 2009.
\newblock ISSN 1095-8541.
\newblock \doi{10.1016/j.jtbi.2009.01.027}.
\newblock URL \url{http://www.ncbi.nlm.nih.gov/pubmed/19490860}.

\bibitem[Orth et~al.(2010)Orth, Thiele, and Palsson]{Orth2010}
Jeffrey~D Orth, Ines Thiele, and Bernhard~\O Palsson.
\newblock {What is flux balance analysis?}
\newblock \emph{Nat. Biotechnol.}, 28\penalty0 (3):\penalty0 245--8, March
  2010.
\newblock ISSN 1546-1696.
\newblock \doi{10.1038/nbt.1614}.
\newblock URL \url{http://www.ncbi.nlm.nih.gov/pubmed/20212490}.

\bibitem[Harrison et~al.(2007)Harrison, Papp, P\'{a}l, Oliver, and
  Delneri]{Harrison2007}
Richard Harrison, Bal\'{a}zs Papp, Csaba P\'{a}l, Stephen~G Oliver, and Daniela
  Delneri.
\newblock {Plasticity of genetic interactions in metabolic networks of yeast.}
\newblock \emph{Proc. Natl. Acad. Sci. U. S. A.}, 104\penalty0 (7):\penalty0
  2307--12, February 2007.
\newblock ISSN 0027-8424.
\newblock \doi{10.1073/pnas.0607153104}.
\newblock URL
  \url{http://www.pubmedcentral.nih.gov/articlerender.fcgi?artid=1892960\&tool=pmcentrez\&rendertype=abstract}.

\bibitem[Mo et~al.(2009)Mo, Palsson, and Herrg\aa~rd]{Mo2009}
Monica~L Mo, Bernhard~\O Palsson, and Markus~J Herrg\aa~rd.
\newblock {Connecting extracellular metabolomic measurements to intracellular
  flux states in yeast}.
\newblock \emph{BMC Syst. Biol.}, 3\penalty0 (1):\penalty0 37, 2009.
\newblock ISSN 1752-0509.
\newblock \doi{10.1186/1752-0509-3-37}.
\newblock URL \url{http://www.biomedcentral.com/1752-0509/3/37}.

\bibitem[Jakubowska and Korona(2012)]{Jakubowska2012}
Agata Jakubowska and Ryszard Korona.
\newblock {Epistasis for Growth Rate and Total Metabolic Flux in Yeast}.
\newblock \emph{PLoS One}, 7\penalty0 (3):\penalty0 e33132, March 2012.
\newblock URL \url{http://dx.doi.org/10.1371/journal.pone.0033132}.

\bibitem[He et~al.(2010)He, Qian, Wang, Li, and Zhang]{He2010}
Xionglei He, Wenfeng Qian, Zhi Wang, Ying Li, and Jianzhi Zhang.
\newblock {Prevalent positive epistasis in Escherichia coli and Saccharomyces
  cerevisiae metabolic networks.}
\newblock \emph{Nat. Genet.}, 42\penalty0 (3):\penalty0 272--6, 2010.
\newblock ISSN 1546-1718.
\newblock \doi{10.1038/ng.524}.
\newblock URL \url{http://www.ncbi.nlm.nih.gov/pubmed/20101242}.

\bibitem[Wall et~al.(2005)Wall, Hirsh, Fraser, Kumm, Giaever, Eisen, and
  Feldman]{Wall2005}
Dennis~P Wall, Aaron~E Hirsh, Hunter~B Fraser, Jochen Kumm, Guri Giaever,
  Michael~B Eisen, and Marcus~W Feldman.
\newblock {Functional genomic analysis of the rates of protein evolution}.
\newblock \emph{Proc. Natl. Acad. Sci. United States Am.}, 102\penalty0
  (15):\penalty0 5483--5488, April 2005.
\newblock \doi{10.1073/pnas.0501761102}.
\newblock URL \url{http://www.pnas.org/content/102/15/5483.abstract}.

\bibitem[Lehner(2011)]{Lehner2011}
Ben Lehner.
\newblock {Molecular mechanisms of epistasis within and between genes}.
\newblock \emph{Trends Genet.}, 27\penalty0 (8):\penalty0 323--331, July 2011.
\newblock \doi{10.1016/j.tig.2011.05.007}.
\newblock URL
  \url{http://www.cell.com/trends/genetics/abstract/S0168-9525(11)00077-1}.

\bibitem[Agrawal and Whitlock(2010)]{Agrawal2010}
Aneil~F Agrawal and Michael~C Whitlock.
\newblock {Environmental duress and epistasis: how does stress affect the
  strength of selection on new mutations?}
\newblock \emph{Trends Ecol. Evol.}, 25\penalty0 (8):\penalty0 450--8, August
  2010.
\newblock ISSN 0169-5347.
\newblock \doi{10.1016/j.tree.2010.05.003}.
\newblock URL \url{http://www.ncbi.nlm.nih.gov/pubmed/20538366}.

\bibitem[Casanueva et~al.(2012)Casanueva, Burga, and Lehner]{Casanueva2012}
M~Olivia Casanueva, Alejandro Burga, and Ben Lehner.
\newblock {Fitness Trade-Offs and Environmentally Induced Mutation Buffering in
  Isogenic C. elegans}.
\newblock \emph{Sci.}, 335\penalty0 (6064):\penalty0 82--85, January 2012.
\newblock \doi{10.1126/science.1213491}.
\newblock URL \url{http://www.sciencemag.org/content/335/6064/82.abstract}.

\bibitem[Frezza et~al.(2011)Frezza, Zheng, Folger, Rajagopalan, MacKenzie,
  Jerby, Micaroni, Chaneton, Adam, Hedley, Kalna, Tomlinson, Pollard, Watson,
  Deberardinis, Shlomi, Ruppin, and Gottlieb]{Frezza2011}
Christian Frezza, Liang Zheng, Ori Folger, Kartik~N Rajagopalan, Elaine~D
  MacKenzie, Livnat Jerby, Massimo Micaroni, Barbara Chaneton, Julie Adam, Ann
  Hedley, Gabriela Kalna, Ian P~M Tomlinson, Patrick~J Pollard, Dave~G Watson,
  Ralph~J Deberardinis, Tomer Shlomi, Eytan Ruppin, and Eyal Gottlieb.
\newblock {Haem oxygenase is synthetically lethal with the tumour suppressor
  fumarate hydratase.}
\newblock \emph{Nature}, 477\penalty0 (7363):\penalty0 225--8, September 2011.
\newblock ISSN 1476-4687.
\newblock \doi{10.1038/nature10363}.
\newblock URL \url{http://www.ncbi.nlm.nih.gov/pubmed/21849978}.

\bibitem[Burgard et~al.(2003)Burgard, Pharkya, and Maranas]{Burgard2003}
Anthony~P Burgard, Priti Pharkya, and Costas~D Maranas.
\newblock {Optknock: A bilevel programming framework for identifying gene
  knockout strategies for microbial strain optimization}.
\newblock \emph{Biotechnol. Bioeng.}, 84\penalty0 (6):\penalty0 647--657,
  December 2003.
\newblock ISSN 1097-0290.
\newblock \doi{10.1002/bit.10803}.
\newblock URL \url{http://dx.doi.org/10.1002/bit.10803}.

\bibitem[Patil et~al.(2005)Patil, Rocha, Forster, and Nielsen]{Patil2005}
Kiran Patil, Isabel Rocha, Jochen Forster, and Jens Nielsen.
\newblock {Evolutionary programming as a platform for in silico metabolic
  engineering}.
\newblock \emph{BMC Bioinformatics}, 6\penalty0 (1):\penalty0 308, 2005.
\newblock ISSN 1471-2105.
\newblock URL \url{http://www.biomedcentral.com/1471-2105/6/308}.

\bibitem[Cornelius et~al.(2011)Cornelius, Kath, and Motter]{Cornelius2011}
Sean~P. Cornelius, William~L. Kath, and Adilson~E. Motter.
\newblock {Controlling Complex Networks with Compensatory Perturbations}.
\newblock May 2011.
\newblock URL \url{http://arxiv.org/abs/1105.3726}.

\bibitem[Covert et~al.(2001)Covert, Schilling, and Palsson]{Covert2001}
Markus~W Covert, C~H Schilling, and Bernhard~\O Palsson.
\newblock {Regulation of gene expression in flux balance models of metabolism.}
\newblock \emph{J. Theor. Biol.}, 213\penalty0 (1):\penalty0 73--88, 2001.
\newblock ISSN 0022-5193.
\newblock \doi{10.1006/jtbi.2001.2405}.
\newblock URL \url{http://www.ncbi.nlm.nih.gov/pubmed/11708855}.

\bibitem[Szappanos et~al.(2011)Szappanos, Kov\'{a}cs, Szamecz, Honti, Costanzo,
  Baryshnikova, Gelius-Dietrich, Lercher, Jelasity, Myers, Andrews, Boone,
  Oliver, P\'{a}l, and Papp]{Szappanos2011}
Bal\'{a}zs Szappanos, K\'{a}roly Kov\'{a}cs, B\'{e}la Szamecz, Frantisek Honti,
  Michael Costanzo, Anastasia Baryshnikova, Gabriel Gelius-Dietrich, Martin~J
  Lercher, M\'{a}rk Jelasity, Chad~L Myers, Brenda~J Andrews, Charles Boone,
  Stephen~G Oliver, Csaba P\'{a}l, and Bal\'{a}zs Papp.
\newblock {An integrated approach to characterize genetic interaction networks
  in yeast metabolism.}
\newblock \emph{Nat. Genet.}, 43\penalty0 (May):\penalty0 656--662, May 2011.
\newblock ISSN 1546-1718.
\newblock \doi{10.1038/ng.846}.
\newblock URL \url{http://www.ncbi.nlm.nih.gov/pubmed/21623372}.

\bibitem[Varma and Palsson(1994)]{Varma1994}
A~Varma and Bernhard~\O Palsson.
\newblock {Stoichiometric flux balance models quantitatively predict growth and
  metabolic by-product secretion in wild-type Escherichia coli W3110.}
\newblock \emph{Appl. Environ. Microbiol.}, 60\penalty0 (10):\penalty0
  3724--31, October 1994.
\newblock ISSN 0099-2240.
\newblock URL
  \url{http://www.pubmedcentral.nih.gov/articlerender.fcgi?artid=201879\&tool=pmcentrez\&rendertype=abstract}.

\bibitem[Brochado et~al.(2012)Brochado, Andrejev, Maranas, and
  Patil]{Brochado2012}
Ana~Rita Brochado, Sergej Andrejev, Costas~D Maranas, and Kiran~R Patil.
\newblock {Impact of Stoichiometry Representation on Simulation of
  Genotype-Phenotype Relationships in Metabolic Networks}.
\newblock \emph{PLoS Comput Biol}, 8\penalty0 (11):\penalty0 e1002758, November
  2012.
\newblock URL \url{http://dx.doi.org/10.1371\%2Fjournal.pcbi.1002758}.

\bibitem[Schellenberger et~al.(2011)Schellenberger, Que, Fleming, Thiele, Orth,
  Feist, Zielinski, Bordbar, Lewis, Rahmanian, Kang, Hyduke, and
  Palsson]{Schellenberger2011a}
Jan Schellenberger, Richard Que, Ronan M~T Fleming, Ines Thiele, Jeffrey~D
  Orth, Adam~M Feist, Daniel~C Zielinski, Aarash Bordbar, Nathan~E Lewis,
  Sorena Rahmanian, Joseph Kang, Daniel~R Hyduke, and Bernhard~O Palsson.
\newblock {Quantitative prediction of cellular metabolism with constraint-based
  models: the COBRA Toolbox v2.0}.
\newblock \emph{Nat. Protoc.}, 6\penalty0 (9):\penalty0 1290--1307, September
  2011.
\newblock ISSN 1754-2189.
\newblock URL \url{http://dx.doi.org/10.1038/nprot.2011.308
  http://www.nature.com/nprot/journal/v6/n9/abs/nprot.2011.308.html\#supplementary-information}.

\bibitem[Barabasi and Oltvai(2004)]{Barabasi2004}
Albert-Laszlo Barabasi and Zoltan~N Oltvai.
\newblock {Network biology: understanding the cell's functional organization}.
\newblock \emph{Nat Rev Genet}, 5\penalty0 (2):\penalty0 101--113, February
  2004.
\newblock ISSN 1471-0056.
\newblock URL \url{http://dx.doi.org/10.1038/nrg1272}.

\end{thebibliography}
